\documentclass[10pt,leqno]{article}
\usepackage{graphicx}
\usepackage{float}
\usepackage{indentfirst,csquotes}

\usepackage{amssymb,amsthm,amsmath}
\usepackage{xcolor,paralist,hyperref,fancyhdr,etoolbox}
\usepackage{algorithm}
\usepackage{algpseudocode}
\usepackage{multirow}
\usepackage{subcaption}
\newtheorem{theorem}{Theorem}[section]

\newtheorem{lemma}[theorem]{Lemma}
\newtheorem{proposition}[theorem]{Proposition}
\newtheorem{corollary}[theorem]{Corollary}

\hypersetup{ colorlinks=true, linkcolor=black, filecolor=black, urlcolor=black }

\usepackage{lipsum}
\usepackage{cleveref}
\begin{document}
\title{A Goal-Oriented Bayesian Experimental Design to Reduce Uncertainty in Sparse-Angle X-ray Computed Tomography}

\author{
Mansoure Giahi Sabour\thanks{University of Oulu, Finland. Email: mansoure.giahisabour@oulu.fi}
\and
Hassan Yazdanian
\and
Babak Maboudi Afkham
}

\date{\today}

\maketitle

\begin{abstract}
Sparse-angle X-ray computed tomography (CT) presents a fundamental challenge for reliable inference, as limited measurements may leave important features of the unknown object, such as its boundaries, unresolved. Existing Bayesian optimal experimental design (BOED) methods provide a principled framework for selecting informative measurements, but can be computationally expensive or require data-driven surrogate models. Moreover, these methods do not explicitly exploit the microlocal visibility of object boundaries. We address this limitation from a goal-oriented perspective, focusing on estimating inclusion boundaries rather than reconstructing the entire attenuation field. For piecewise-constant objects considered in this study, these boundaries correspond to singularities whose locations and directions are characterized by their wavefront sets. Microlocal analysis determines the X-ray view-angles at which these singularities are visible. We combine this information with Bayesian uncertainty quantification to introduce B-MUSE, a microlocal uncertainty-driven sequential experimental design method. Specifically, we define a microlocal C-optimality criterion associated with the wavefront direction of the most uncertain boundary singularity and prove that its optimal solution corresponds to a measurement satisfying the microlocal visibility condition. B-MUSE uses this result to sequentially select informative view-angles according to posterior boundary uncertainty. It also incorporates an adaptive stopping criterion that terminates acquisition when additional measurements provide no meaningful reduction in uncertainty. We evaluate B-MUSE through numerical experiments of increasing complexity, including analytically interpretable examples, inclusions generated from the assumed prior, and a multi-inclusion example from an external dataset. We further benchmark B-MUSE against recent experimental design methods. The results show that B-MUSE substantially reduces posterior uncertainty with a small number of adaptively selected measurements while providing accurate boundary estimates. These findings establish a direct connection between BOED and microlocal visibility, demonstrating the potential of uncertainty-driven acquisition to reduce measurement requirements for goal-oriented X-ray CT without the need for training data.
\end{abstract} 
\bigskip
\section{Introduction}
X-ray computed tomography (CT) is an imaging technique whose applications span a wide range of fields, including industrial inspection, security screening, archaeology, and medical imaging \cite{oliveira2019characterization, ParkerSecurityCT, de2024revealing, hansen2021computed, kaipio2006statistical}. It can be formulated as an \emph{inverse problem}, i.e., reconstruction of the interior of an object from indirect X-ray attenuation measurements acquired outside the object. In many X-ray CT applications, practical considerations such as acquisition time, radiation exposure, and physical constraints of the imaging setup can limit the number of available measurements \cite{TsengBreastCBCT,Hatamikia2021CarmCBCT,Hyvarinen2026CBCT}. In the absence of an adequate amount of measurement data, e.g., in the presence of data sparsity, X-ray CT is ill-posed. Several different interior configurations may be consistent with similar measurement patterns \cite{hansen2021computed}. In such cases, obtaining a reliable and accurate reconstruction of the entire object becomes more challenging.

When the ultimate goal of X-ray CT is to extract low-dimensional features or quantities of interest (QoIs) from the imaged object, data sparsity can lead to reconstruction artifacts that compromise the accuracy of the extracted information. Recent studies \cite{afkham2024uncertainty,BabakUQinclusionB}, however, suggest that this limitation can be mitigated by reformulating the inverse problem to infer the desired QoIs directly from the X-ray measurements, rather than first reconstructing the entire image. Under this \emph{goal-oriented} inverse formulation, the number of X-ray measurements required to reliably estimate the QoIs can be substantially reduced. This observation raises a natural question: if only a limited number of X-ray measurements are required, how should these measurements be optimally selected to maximize the information relevant to the QoIs?

A representative example of this setting arises in tumor or target volume contouring in radiation oncology \cite{vinod2016uncertainties}, where target valume boundaries are delineated from reconstructed X-ray CT images, for example in breast-cancer radiotherapy \cite{BreastSegmentationReview, TargetContouring}. In conventional imaging workflows, the reliability of such delineations depends strongly on reconstruction quality, and artifacts caused by sparse or insufficient X-ray data may lead to inaccurate boundary estimation and, consequently, affect treatment planning. This application motivates the present work: rather than seeking sufficient measurements for accurate reconstruction of the entire image, we aim to identify an optimal subset of X-ray measurements that enables reliable estimation of the boundaries of inclusion.

A common choice to mathematically represent such inclusions is through piecewise-constant models consisting of approximately homogeneous regions \cite{piecewise1, piecewise2, piecewise3}. The boundaries between these regions correspond to discontinuities or singularities of the unknown object. Microlocal analysis \cite{krishnanQuintoMicrolocal} provides a mathematical framework for identifying, characterizing, and classifying such singularities and for determining whether these singularities are captured by the measured data \cite{Quint-visibility}.
Sparse measurements may fail to capture the information required to stably reconstruct some of these singularities.
The implications of this phenomenon for filtered back projection (FBP) \cite{hansen2021computed}, a conventional direct reconstruction method, are examined in \cite{FrikelXray_Artfct,borgQuinto}. These studies show that certain true singularities cannot be stably reconstructed, while additional artifacts may appear in the reconstructed image that are not present in the original object.

Recent works have also explored incorporating microlocal information directly into the reconstruction process. In \cite{FRIKELSpars}, the curvelet coefficients of the unknown object are divided into visible and invisible components according to the available measurements. The reconstruction is then formulated using sparse regularization over only the visible curvelet coefficients, 
leading to an edge-preserving reconstruction with a reduced-dimensional formulation. The idea of separating visible and invisible features was later incorporated in learned reconstruction methods. In \cite{learn-invisible}, a model-based sparse regularization reconstruction is first used to recover the visible shearlet coefficients from the available measurements. These coefficients are retained, while the invisible coefficients, which cannot be stably recovered from data, are inferred by a neural network. The visible and inferred invisible sets of shearlet coefficients are then combined and transformed back to the image domain.

A more boundary-oriented approach is proposed in \cite{Siiri_wf_learnPrior}.  Visible boundary information is first extracted from a reconstruction using wavelets and a learned model. This information is then extended to the invisible regions using a model for the expected continuation of the boundary. A second neural network predicts the complete singular support, which is subsequently used for boundary reconstruction. In \cite{DeepMicrolocal}, microlocal information is incorporated more directly by jointly learning the image and its wavefront set, which encodes singularity locations and directions. The visible wavefront information is first extracted from the measurements and then mapped into the image domain, after which a neural network is used to infer the remaining invisible wavefront information. 

The methods discussed above are deterministic approaches that assume a fixed acquisition geometry and exploit microlocal information available in the X-ray measurements to improve reconstruction. In sparse-data settings, however, the measurements may not contain sufficient information to determine all features of the unknown, making the choice of acquisition geometry itself an important part of the inverse problem. The Bayesian framework \cite{kaipio2006statistical} provides a probabilistic formulation for a given acquisition geometry by combining the information provided by the measurements, through the \emph{likelihood}, with prior knowledge of the unknown, encoded by the \emph{prior} distribution. Combining the prior and likelihood distributions, through the Bayes' theorem, yields the \emph{posterior}  distribution, which characterizes plausible reconstructions conditioned on the observed data.
Qualities of the posterior can be interpreted as uncertainty in reconstruction, highlighting credible singularities.


This idea naturally leads to Bayesian optimal experimental design (BOED) for X-ray CT. In BOED, candidate measurement configurations are evaluated according to the posterior uncertainty they are expected to produce before the measurements are acquired. The uncertainty can be summarized through posterior moments, such as the covariance, and used to define different optimality criteria
\cite{Attia_2018, orozcoBOED,alexanderian_Review_InfD}. Common choices include A-, D-, and C-optimality. A-optimality seeks to minimize the trace of the posterior covariance, D-optimality seeks to minimize the log-determinant of the posterior covariance, and C-optimality focuses on minimizing the posterior uncertainty in a specified linear functional of the unknown.  

Within this framework, several sequential experimental design methods have been developed for X-ray CT. A greedy sequential algorithm for selecting an informative set of measurements is proposed in \cite{seq-optim-Xray}. The authors consider a linear Gaussian model, in which the posterior covariance obtained from previous measurements serves as the prior when selecting the next measurements. Both A- and D-optimality criteria are evaluated, with design variables related to the X-ray CT geometry. Another greedy sequential method is introduced in \cite{helinXrayBED}. There, the posterior is approximated by a Gaussian distribution whose covariance incorporates information about the locations of edges in the reconstructed image. The next measurement is then selected according to either A- or D-optimality criterion. In both works, the posterior covariance is tractable, either exactly under the Gaussian model in the first work or through a Gaussian approximation in the second, and can therefore be used to guide the acquisition. For more general non-Gaussian models, however, evaluating such covariance-based optimality criteria can be computationally expensive. To address this difficulty, \cite{PACE}, A-optimal design is reformulated to avoid repeated evaluations of posterior covariance. A Neural network is instead trained to estimate the conditional expectation of the unknown given the measurements associated with a candidate design. 

Despite these developments, classical BOED approaches can be computationally prohibitive, as the resulting optimization problem often involves expectations with respect to an intractable posterior distribution. Data-driven approaches \cite{PACE,Node} can alleviate the need for repeated posterior computations, but typically require representative training samples together with the construction and training of neural networks or other surrogate models. Consequently, the computational cost of posterior-based design, either through direct Bayesian inference or its data-driven approximation, remains a significant challenge for optimal measurement selection.

Beyond computational cost, the BOED methods discussed above aim to select informative measurements, but they do not exploit microlocal analysis, which can characterize the visibility of singularities under a given acquisition geometry. 
Bayesian approaches incorporating microlocal analysis have also been studied \cite{microBayes1, MicroBayes2}, although with a different purpose. These works provide uncertainty quantification (UQ), but they do not use the resulting uncertainty to guide data acquisition or to solve an experimental design problem aimed at reducing it.
In \cite{microBayes1}, microlocal analysis is used to investigate posterior behavior, whereas \cite{MicroBayes2} studies infinite-dimensional Bayesian inference for X-ray tomography from a microlocal perspective.

Motivated by these limitations, in this paper, we consider the reconstruction of inclusion boundaries in sparse-angle X-ray CT. We formulate the problem in a Bayesian framework and use a star-shaped prior distribution for piecewise-constant images to parameterize the inclusion boundaries. Combining this prior with the likelihood associated with the available sparse measurements yields a posterior distribution that characterizes the uncertainty in the boundary locations. A central hypothesis of this work is that the optimal measurement angles are those that maximally reduce the posterior uncertainty in the inclusion boundaries.

To this end, we combine microlocal analysis and BOED. In particular, we define a microlocal C-optimality criterion based on a linear functional associated with the wavefront direction of the most uncertain singularity. We show that, under the stated assumptions, the solution of this optimization problem corresponds to the measurement satisfying the microlocal visibility condition. 
Based on this result, we propose the \emph{B-MUSE} algorithm, a Bayesian microlocal uncertainty-driven sequential experimental design method that uses the proposed microlocal C-optimality criterion to select the most informative measurement. We also equip B-MUSE with an adaptive stopping criterion that determines when additional measurements no longer provide a meaningful reduction in uncertainty. We evaluate the performance of B-MUSE through a series of experiments of increasing complexity, beginning with a setting in which its behavior can be understood analytically, followed by a more complex inclusion generated from the prior, and ultimately a multi-inclusion example not directly sampled from the prior distribution. To further assess its performance, we benchmark B-MUSE against recent experimental design methods.  

The significance of B-MUSE lies in combining Bayesian UQ and microlocal analysis, and using both directly to guide the acquisition process. To the best of our knowledge, this is the first work to establish an analytical connection between BOED and microlocal analysis for sequential view-angle selection in X-ray CT. Rather than relying on a fixed acquisition geometry, B-MUSE adapts subsequent measurements according to the posterior uncertainty in the inclusion boundaries. Moreover, B-MUSE is a model-driven method and does not require a training dataset or a learned surrogate model. Our numerical experiments show that this sequential design can substantially reduce posterior uncertainty with relatively few X-ray projections. 
Although we illustrate the proposed method using piecewise-constant attenuation fields, the underlying theory is not restricted to this setting and can naturally be extended to piecewise-smooth models, provided that the prior encodes the corresponding wavefront information.

The remainder of this paper is organized as follows. In \Cref{sec:MathCT}, we introduce the mathematical formulation of sparse-angle X-ray CT and its forward model. In \Cref{sec:BayesForm} we present the Bayesian formulation for this problem. In \Cref{sec:microlocal-analysis}, we review the microlocal analysis of X-ray CT and the associated visibility conditions. In \Cref{sec:BOED}, we review BOED criteria. In \Cref{sec:adaptive-angle-selection}, we define the microlocal C-optimality criterion and introduce the B-MUSE algorithm. In  \Cref{sec:Numerical-r}, we present the numerical experiments, including the disk and prior-sample experiments, comparisons with other experimental design methods, and finally the multi-inclusion experiment based on an external dataset. The paper concludes in \Cref{sec:conclusion} with a summary of the main findings, limitations, and directions for future work.


\section{Mathematical model of sparse-angle X-ray CT} \label{sec:MathCT}
 X-ray CT is an inverse problem, since the goal is to reconstruct the internal structure of an object from external X-ray measurements.  In this section, we first review the physical measurement process of the X-ray CT and then introduce the forward operator that links this physics to a mathematical model. Finally, we present its mathematical formulation used throughout this paper.

In X-ray CT, an object is probed by X-rays from multiple directions. We refer to the direction in which X-rays propagate through the object as a \emph{view-angle}. At each view-angle, the X-ray intensity is attenuated according to the attenuation properties of the materials along its propagation path \cite{hansen2021computed,kaipio2006statistical}, and the attenuated intensities are then recorded by a detector. The collection of all measurements associated with different view-angles forms a \emph{sinogram}, from which information about the internal structure of the object can be reconstructed. 

To formulate the reconstruction problem mathematically, the physical attenuation process is represented by a \emph{forward operator}, which maps  a known attenuation field to the corresponding measurements. In the parallel-beam setting considered here,  this forward operator is modeled by the \emph{Radon Transform}, in which each measurement is represented as a line integral of the attenuation field.

Let the attenuation field be a function $x \in L^2(\Omega)$, defined on a bounded domain $\Omega \subset \mathbb{R}^2$. For a view-angle \(\theta\in[0,\pi),\) define
\( \boldsymbol{\theta}= \begin{pmatrix}
\cos\theta& \sin\theta \end{pmatrix}^T\) 
 and \( \boldsymbol{\theta}^{\perp} =\begin{pmatrix}
-\sin\theta& \cos\theta \end{pmatrix}^T.
\) Here, $\boldsymbol{\theta}$ denotes the X-ray propagation direction, and $\boldsymbol{\theta}^\perp$ is the corresponding normal direction. 
For a detector coordinate \(s\in S\subset\mathbb R\), the corresponding X-ray line is $\ell_{\theta,s} = \{ \xi \in \mathbb{R}^2 : \xi \cdot \boldsymbol{\theta}^\perp = s \}.$ The continuous Radon transform is then given by \cite{hansen2021computed}
\begin{equation} \label{eq:cts-Radon}
\mathcal{R}[x](\theta, s) = \int_{\ell_{\theta,s}} x(\xi)\, d\ell,
\end{equation}
where $d\ell$ is the line element along \(\ell_{\theta,s}\), $\xi \in \mathbb R^2$ is the spatial variable, and \(s\) is the signed distance of the integration line
from the origin. Thus, the Radon transform maps the attenuation field
to its line integrals. With 
\emph{complete angular data}, i.e., measurements over the
full range of view-angles \(\theta\in[0,\pi)\), the Radon transform
uniquely determines \(x\)
\cite{hansen2021computed,kaipio2006statistical}.

In practice, the continuous Radon transform in  \eqref{eq:cts-Radon} is evaluated only at a finite number of view-angles and detector locations.
Let \(\Theta=\{\theta_i\}_{i=0}^{N_{\mathrm view}-1} \subset [0,\pi) \) denote the set of view-angles.
 Let \(S_{detect}=\{s_q\}_{q=1}^{N_{\mathrm{detect}}}
\subset S\) denote the detector locations. For each $\theta_i \in \Theta, \ i=0,\ldots,N_{\mathrm{view}}-1$, we define the single view-angle measurement operator \(\mathcal{R}_{\theta_i}: L^2(\Omega) \to \mathbb R^{ N_{\mathrm detect}} \) by the rule
\begin{equation} \label{eq:single-angle-R}
    \bigl[\mathcal R_{\theta_i}x\bigr]_{q}
=
\mathcal R[x](\theta_i,s_q),
\quad
q=1,\ldots,N_{\mathrm{detect}}.
\end{equation}   We stack the projection measurements over all acquired view-angles and define the total measurement operator \(  \mathcal R_{\Theta}: L^2(\Omega) \to \mathbb R^{N_{\mathrm{sino}}},\) where $N_{\mathrm{sino}}
=
N_{\mathrm{view}} \times N_{\mathrm{detect}}$ with the rule 
\begin{equation}\label{eq:finite-total-radon}
     \  \mathcal R_{\Theta}x
    =
    \bigl(
        (\mathcal R_{\theta_0}x)^{T},
        \ldots,
        (\mathcal R_{\theta_{N_{\mathrm{view}}-1}}x)^{T}
    \bigr)^{T}.
\end{equation}  
Thus, the continuous attenuation field is mapped to a finite-dimensional vector of measurements. When available measurements are restricted to only a small number of view-angles, the problem is referred to as \emph{sparse-angle tomography}. 

We consider an additive Gaussian noise model and write the observed
data as
\begin{equation}
\label{eq:additive-noise-model}
\boldsymbol y_{\Theta}
=
\mathcal R_{\Theta}x
+
\boldsymbol{\varepsilon}_{\Theta},
\qquad
\boldsymbol{\varepsilon}_{\Theta}
\sim
\mathcal N
\left(
0,\sigma^2 I_{N_{\mathrm{sino}}}
\right).
\end{equation}
Here,
\(\boldsymbol y_{\Theta}\in\mathbb R^{N_{\mathrm{sino}}}\) is the observed
sinogram vector, whereas
\(\mathcal R_{\Theta}x\) is the corresponding noise-free
measurement vector.

In tomographic problems, noise is rarely strictly Gaussian, and the assumption of a Poisson-distributed noise is more realistic. However, under sufficiently high photon intensity in the detector, the Poisson distribution admits a Gaussian approximation through the central limit theorem, which is often simpler to work with \cite{hansen2021computed, kaipio2006statistical}.
Moreover, the Gaussian noise assumption leads to a more tractable probabilistic formulation. In the next section, we introduce the Bayesian formulation, where the unknown, measurements, and noise are modeled as random variables.
\section{Bayesian formulation}\label{sec:BayesForm}
In \Cref{sec:MathCT}, we related the attenuation field $x$ to the finite-dimensional noisy measurement vector $\boldsymbol y_\Theta$ through  \eqref{eq:additive-noise-model}.  In sparse-angle tomography, the missing view-angles may introduce uncertainty in the reconstructed image.
To quantify this uncertainty, we 
formulate the observation model in  \eqref{eq:additive-noise-model} probabilistically by expressing the unknown attenuation field, noise, and measurements as random variables. The Bayesian solution of the inverse problem in this setting is called the posterior probability distribution, i.e., the distribution of the unknown attenuation field conditioned on the observed measurements. We can formally apply Bayes' rule to characterize this posterior distribution by combining prior information about the unknown before observing the measurements with the likelihood, which is the probability of the observed measurements for a given attenuation field. To construct these components, we first review Gaussian random variables on Hilbert spaces and their Karhunen-Lo\`eve expansion.

\subsection{Gaussian random variables on Hilbert spaces} \label{ssec:H-gaussian-random-var}

Let $(\Gamma_{\mathrm{prob}},\mathcal F,\mathcal P)$ be a probability space and let $(\mathcal H,\langle\cdot,\cdot\rangle_{\mathcal H})$ be a separable Hilbert space. An $\mathcal H$-valued random variable is a measurable mapping $Z:\Gamma_{\mathrm{prob}}\to\mathcal H$. The random variable $Z$ is called Gaussian if $\langle Z,g\rangle_{\mathcal H}$ is a real-valued Gaussian random variable for every $g\in\mathcal H$ \cite{dashti2017bayesian,stuart2010inverse}. A Gaussian random variable is characterized by its mean $m\in\mathcal H$ and covariance operator $\mathcal C:\mathcal H\to\mathcal H$, where $\mathcal C$ is self-adjoint, positive semidefinite, and trace class \cite{Ibragimov1978Gaussian}. We denote the corresponding Gaussian measure by $\mu=\mathcal N(m,\mathcal C)$.

Assume a set of eigenfunctions $\{e_k\}_{k\in\mathbb N}$  and nonnegative eigenvalues $\{\lambda_k\}_{k \in \mathbb N }$ of $\mathcal C$, sorted in non-increasing order, such that \(\mathcal C e_k = \lambda_k e_k.\) It is shown in \cite{Ibragimov1978Gaussian} that $Z$ admits a \emph{Karhunen-Lo\'eve} expansion 
 \begin{equation} \label{eq:KL}
     Z= \sum_{k \in \mathbb N} \sqrt{\lambda_k} Z_k e_k +m,
 \end{equation}
where $\{Z_k\}_{k\in\mathbb N}$ is  a sequence of mutually independent, real-valued random variables with the distribution \(\mathcal N(0,1)\). 

\subsection{Prior modeling} \label{ssec:prior}
In many imaging applications, the unknown attenuation field contains distinct regions in which the attenuation coefficient is approximately constant \cite{piecewise1, piecewise2, piecewise3}.   Under this assumption, the attenuation field can be represented as the  $L^2(\Omega)$-valued random variable \(  X:\Gamma_{\mathrm{prob}}\to L^2(\Omega) \) with piecewise-constant realizations. One way to define such random images is through their realizations. Suppose that $x$ is a realization of $X$. Then we express $x$ as
\begin{equation} \label{eq:Xpiecewise}
    x(\xi)=
\begin{cases}
\alpha_d^+, & \xi\in D_d,\quad d=1,\dots,N_{inc},\\
\alpha^-, & \xi\notin \bigcup_{d=1}^{N_{inc}} D_d,
\end{cases}
\end{equation}
where $\xi\in \Omega$ is the spatial coordinate, $D_d \subset \Omega,\ d=1,\ldots,N_{inc} $ are closed bounded regions with Lipschitz continuous boundaries, and 
\(   \alpha^-,\alpha_d^+\in\mathbb R \) denote the attenuation coefficients of the background and inside each  \(D_d\), respectively. Assuming $\alpha^-$ and $  \{\alpha_d^+\}_{d=1}^{N_{\mathrm{inc}}} $ are known, there is an immediate relation between the boundaries of $D_d$ and $x$ through \eqref{eq:Xpiecewise}. In other words, knowing the boundaries of $D_d$ uniquely determines the image $x$, and the unknown information is entirely contained in the boundaries of $D=\bigcup_{d=1}^{{N_{inc}}}D_d.$ Therefore, rather than reconstructing the entire attenuation field $x,$ we focus on reconstructing the boundaries in $D,$ which leads to a lower dimensional inverse problem. 
\begin{figure}[H]\centering\includegraphics[width=0.35\linewidth]{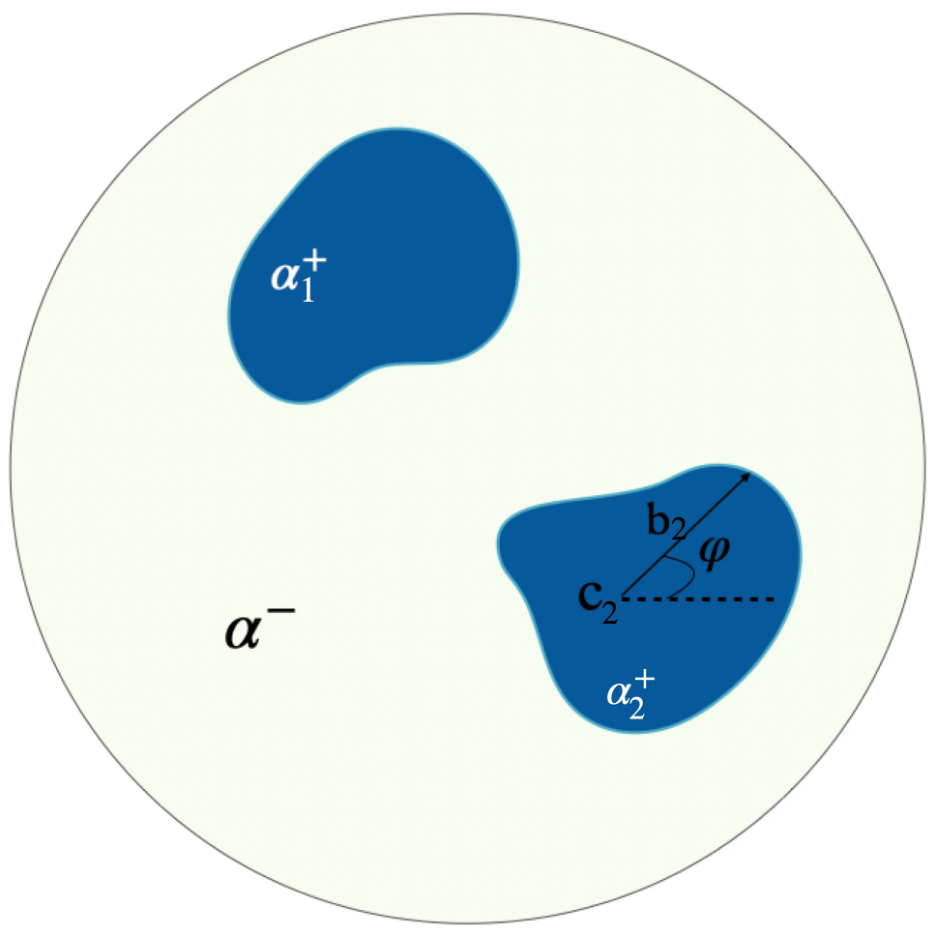}
    \caption{One attenuation field with two star-shaped inclusions.  Inclusion are identified by  centers $c_d$ and radial functions $b_d(\varphi),\ \varphi \in [0,2\pi),~ d=1,2$.}
    \label{fig:starshaped}
\end{figure}

To model the boundaries in $D_d, \ d=1,\dots, N_{inc}$, we assume that $D$ comprises disjoint \emph{star-shaped} objects \cite{BabakUQinclusionB}. Such objects can be described by a center random variable $C_d: \Gamma_{prob} \to  \Omega$ together with a random, and periodic, radial function $B_d:\Gamma_{\text{prob}}\to L^2([0,2\pi))$. For each realization $c_d$ and $b_d$, the corresponding radius $b_d(\varphi),$ $\varphi \in [0,2\pi),$ determines the distance from the center $c_d$ to the boundary $D_d$ in all angular directions.  We present a visualization of the domain with such star-shaped inclusions in Figure  \ref{fig:starshaped}. 

For modeling the radial function of each inclusion $D_d, \ d=1,\dots,N_{inc}$, we introduce a latent Gaussian   $L^2([0,2\pi))$-valued random variable  $Z_d:\Gamma_{\text{prob}} \to L^2([0,2\pi))$. We then define a positive radial function through the transformation \(B_d(\varphi):=b_0 + s_B\exp(Z_d(\varphi)),\) where $b_0\geq 0$ is the base radius and $s_B>0$ controls the amplitude of the radial function. Since the radial function $B_d$ is obtained from the latent variable $Z_d$, for \( C=(C_1,\dots,C_{N_{inc}}) \) and \( Z=(Z_1,\dots,Z_{N_{inc}}) \) the collection of corresponding star-shaped objects are determined by the pair $(C,Z):\Gamma_{\text{prob}}\to \Omega ^{N_{inc}} \times \big(L^2([0,2\pi])\big)^{N_{inc}} $. Under this parameterization, the attenuation field in \eqref{eq:Xpiecewise} is represented through the mapping \( F_H(C,Z):  \Gamma_{\text{prob}} \to L^2(\Omega)\), defined by
\begin{equation}\label{eq:F-Heaviside}
F_H(C,Z)(\xi)= \alpha^- +\sum_{d=1}^{ N_{\mathrm{inc}}}  (\alpha_d^+-\alpha^- )\,K_H\!\left(B_d(\varphi)-\|\xi-C_d\|\right), 
\qquad
K_H(t)=
\begin{cases}
1, & t\ge 0,\\
0, & t<0.
\end{cases}
\end{equation}
where $K_H$ denotes the Heaviside function \cite{mathFuncBook}. In this representation, the condition \( B_d(\varphi)-\|\xi-C_d\|=0\) determines the boundaries of the $d$th region for all $d\leq N_{\mathrm{inc}}.$ Although this parametrization is useful, the mapping $F_H(C,Z)$ is non-differentiable with respect to $C$ and $Z$. Therefore, we replace the Heaviside function with a sigmoid approximation, yielding the smoothed attenuation field 
\begin{equation}\label{eq:F-Sigmoid}
F_{S}(C,Z)(\xi)= \alpha^- +\sum_{d=1}^{N_{\mathrm{inc}}}  (\alpha_d^+-\alpha^- )\,K_{S}\!\left(B_d(\varphi)-\|\xi-C_d\|\right), \qquad K_S(t)=\frac{1}{1+\exp{(-\beta t)}}, 
\end{equation} 
where $\beta >0$ controls the sharpness of the transition across the boundary. As $\beta \rightarrow \infty$, the transition becomes increasingly sharp and $F_S(C,Z)\rightarrow F_H(C,Z)$  in $L^2[\Omega] $. In the next sections, we use the smooth mapping $F_S(C,Z)$.

We now describe the construction of a prior distribution for the joint random variable $(C,Z)$. We assume that $C_d$ and $Z_d$ are independent. 
For each $d=1,\ldots,N_{\mathrm{inc}}$,
we let $C_d\sim\mu_{C_d}:=\mathcal U(\Omega)$, where $\mathcal U(\Omega)$ denotes the uniform (Lebesgue) probability measure on $\Omega$. Moreover, we assign to each $Z_d$ an independent Gaussian prior. Recall from \Cref{ssec:H-gaussian-random-var} that a Gaussian random variable is fully characterized by its mean function and covariance operator. In this work, we consider covariance operators from the Mat\'ern family \cite{Matern2013, studun2016}, defined by $\mathcal C_{\mathrm{Mat\acute ern}} := (\kappa^2-\partial^2/\partial \varphi^2)^{-\alpha},$ subject to periodic boundary conditions, where $\kappa>0$ determines the angular correlation length and $\alpha>1$ controls the smoothness of the resulting boundaries. We then define $Z_d\sim\mu_{Z_d}:=\mathcal N(0,\mathcal C_{\mathrm{Mat\acute ern}})$. 
The prior distribution for the joint random variable $(C,Z)$ is then given by the product measure $\mu_{(C,Z)}:=\mu_C\otimes\mu_Z,$
where \( \mu_Z = \otimes_{d=1}^{N_\mathrm{inc}} \mu_{Z_d}  , \   \mu_C = \otimes_{d=1}^{N_\mathrm{inc}} \mu_{C_d} .  \)

Having defined the prior model for the unknown boundary, we now formulate the likelihood associated with the current measurements in the next section. 
\subsection{Noise model, likelihood, and posterior}
\label{noise-Lklhood-post2}

In \Cref{sec:MathCT}, the additive noise model
\eqref{eq:additive-noise-model} was formulated in terms of the
attenuation field \(x\), the observed measurement vector
\(\boldsymbol y_\Theta\), and a realization of the measurement noise
\(\boldsymbol\varepsilon_\Theta\). We now formulate the same observation model in terms of random variables.

For each view-angle
\(\theta_i\in\Theta\), let
\(
\boldsymbol E_{\theta_i}:
\Gamma_{\mathrm{prob}}
\longrightarrow
\mathbb R^{N_{\mathrm{detect}}}
\) denote the corresponding measurement noise vector. We assume that the
noise vectors are mutually independent, and
identically distributed (i.i.d) according to
\begin{equation} \label{noise-model}
 \boldsymbol E_{\theta_i}
\stackrel{\mathrm{i.i.d.}}{\sim}
\mathcal N
\left(
0,\sigma^2 I_{N_{\mathrm{detect}}}
\right),
\
i=0,\ldots,N_{\mathrm{view}}-1,
\end{equation}  
where $\sigma>0$ is noise standard deviation. Here, for simplicity in representation, we choose a constant $\sigma$ for all view-angles. However, in practice, each view-angle may have different noise level. We refer the reader to \Cref{sec:Numerical-r} for further detail.  Let \(\boldsymbol Y_{\theta_i}:
\Gamma_{\mathrm{prob}}\to
\mathbb R^{N_{\mathrm{detect}}}\)
denote the random measurement vector corresponding to the view
angle \(\theta_i\). The statistical single view-angle  measurement model is given by
\begin{equation} \label{stat-single-angle-forward}
    \boldsymbol Y_{\theta_i}
=
\mathcal R_{\theta_i}\bigl(F_S(C,Z)\bigr)
+
\boldsymbol E_{\theta_i},
\qquad
i=0,\ldots,N_{\mathrm{view}}-1.  
\end{equation}
Stacking the single view-angle measurement vectors gives \( \boldsymbol Y_\Theta
= ( \boldsymbol Y_{\theta_0}^{T},  \ldots,
\boldsymbol Y_{\theta_{N_{\mathrm{view}}-1}}^{T} )^{T} \) which results in the statistical forward model
\begin{equation}\label{stat-total-forward-model}
\boldsymbol{Y}_{\Theta}=\mathcal{R}_{\Theta}\bigl(F_S(C,Z)\bigr)+\boldsymbol E_{\Theta},
\end{equation}
where $\boldsymbol E_{\Theta}=\bigl(\boldsymbol E_{\theta_0}^T, \dots \boldsymbol E_{\theta_{N_{\mathrm{view}}-1}}^T\bigr)^T$ is the stacked noise vector. 
Equation \eqref{stat-total-forward-model} is therefore the statistical counterpart of \eqref{eq:additive-noise-model}.

Since the noise vectors are assumed to be independent,  the single view-angle measurement vectors
\(\{\boldsymbol Y_{\theta_i}\}_{i=0}^{N_{\mathrm{view}}-1}\) are conditionally independent given $(C, Z)$. Consequently, the likelihood of the observed measurements factorizes as
\begin{equation}
\label{eq:likelihood}
\begin{aligned}
&\mu_{\{\boldsymbol Y_{\theta_i}\}_{i=0}^{N_{\mathrm{view}}-1}
\mid(C,Z)}
\left(
\boldsymbol{y}_{\theta_0},\ldots,
\boldsymbol{y}_{\theta_{N_{\mathrm{view}}-1}}
\mid \boldsymbol{c},\boldsymbol{z}
\right)
\\
&\quad =
\prod_{i=0}^{N_{\mathrm{view}}-1}
\mu_{\boldsymbol Y_{\theta_i}\mid(C,Z)}
\left(
\boldsymbol{y}_{\theta_i}
\mid\boldsymbol{c},\boldsymbol{z}
\right)
\quad \propto
\prod_{i=0}^{N_{\mathrm{view}}-1}
\exp\!\left(
-\frac{
\left\|
\boldsymbol{y}_{\theta_i}
-
\mathcal R_{\theta_i}
\bigl(F_S(\boldsymbol{c},\boldsymbol{z})\bigr)
\right\|_2^2
}{
2\sigma^2
}
\right).
\end{aligned}
\end{equation}
Given the prior measure $\mu_{(C,Z)}=\mu_C\otimes\mu_Z$, Bayes' theorem yields the posterior measure
$\mu_{(C,Z)\mid \boldsymbol Y_\Theta=\boldsymbol{y}_\Theta}$. Following the framework of \cite{stuart2010inverse}, the posterior can be expressed through the Radon-Nikodym derivative
\begin{equation}\label{eq:posterior}
\frac{
d\mu_{(C,Z)\mid \boldsymbol Y_\Theta=\boldsymbol{y}_\Theta}
}{
d\mu_{(C,Z)}
} (\boldsymbol{c},\boldsymbol{z})
=
\frac{1}{c_{\mathrm{post}}}
\prod_{i=0}^{N_{\mathrm{view}}-1}
\exp\!\left(
-\frac{
\left\|\boldsymbol{y}_{\theta_i}-\mathcal{R}_{\theta_i}\bigl(F_S(\boldsymbol{c},\boldsymbol{z})\bigr)\right\|_2^2
}{
2\sigma^2
}
\right),
\end{equation} 
where $c_{\mathrm{post}}$ is the normalizing constant. It was shown in \cite{BabakUQinclusionB} that the posterior distribution constructed in this manner is well defined, i.e., $0<c_{\mathrm{post}}<\infty$. Furthermore, the posterior depends Lipschitz continuously on the observed data with respect to the Hellinger metric.

\subsection{Uncertainty Quantification and Posterior Sampling } \label{ssec:uq-posterior-sampling}
In the previous section we described how to formulate the goal-oriented posterior distribution for inferring inclusion boundaries, expressed in \eqref{eq:posterior}. In this section we describe how we can explore this distribution to quantify the uncertainties in the estimation.

Recall that a posterior distribution on $(C,Z)$ also induces a distribution on image space through the push-forward notation $\mu_{X \mid \boldsymbol Y_{\Theta}= \boldsymbol y_\Theta} = (F_{S})_\#\mu_{(C,Z)\mid \boldsymbol Y_\Theta=\boldsymbol y_\Theta}$, where $F_{S}$ denotes the parameter-to-image map introduced in \eqref{eq:F-Sigmoid}. Therefore, we can investigate uncertainty both in boundary space, i.e., for $(C,Z)$, and image space.


Let $\Upsilon$ be a generic random variable. The first moment of the posterior distribution is given by the expected value, $\mathbb E\bigl[\Upsilon| \boldsymbol Y_\Theta = \boldsymbol y_\Theta\bigr]$. To quantify reconstruction uncertainty, we consider posterior variance, given by
\begin{equation} \label{eq:post-var}
    \text{Var}(\Upsilon |\boldsymbol Y_\Theta=\boldsymbol y_\Theta) := \mathbb E\big[
\left(
\Upsilon
-
\mathbb E\left[
\Upsilon\mid\boldsymbol Y_\Theta=\boldsymbol y_\Theta
\right]
\right)^2
\;\mid\;
\boldsymbol Y_\Theta=\boldsymbol y_\Theta
\big].
\end{equation}
Since the posterior distribution is known up to a normalization constant and arises from a nonlinear model, its moments do not generally admit closed-form expressions.
 Therefore, in this work, we approximate them using ergodic estimates obtained from posterior samples. To generate these samples,  we employ the No-U-Turn Sampler (NUTS)\cite{hoffman2014nuts}, an adaptive variant of Hamiltonian Monte Carlo (HMC). By exploiting gradient information from
the posterior density, NUTS can mitigate the random-walk behavior of conventional random-walk MCMC methods, enabling more efficient posterior exploration. Moreover, it automatically adapts the trajectory length, eliminating the need to specify it manually.

Given a set of posterior samples $\{ (\boldsymbol{c}_s,\boldsymbol{z}_s) \}_{s=1}^{N_s}$, with $N_s$ denoting the number of samples, we approximate the posterior moments defined above in terms of ergodic averages of the samples as
\begin{equation} \label{eq:ergodic}
    \mathbb E[f(C,Z)] \approx \frac{1}{N_s} \sum_{s=1}^{N_s} f(\boldsymbol{c_s},\boldsymbol{z}_s),
\end{equation}
where $f$ is a real-valued or a vector-valued function.



\section{Microlocal analysis and wavefront sets in X-ray CT} \label{sec:microlocal-analysis}
In the previous section, we introduced a Bayesian framework for quantifying uncertainty in the attenuation field and inclusion boundaries. We now turn to microlocal analysis, which provides a mathematical framework for studying singularities in functions and how they transform under the action of operators \cite{hormander, frieddistributions}.  In X-ray CT, microlocal analysis provides a framework for understanding the singularities in the attenuation field and their relationship to those of the corresponding sinogram through the Radon transform \cite{krishnanQuintoMicrolocal}.

We start by introducing the notion of local singularity.
From a microlocal perspective, smoothness of a function is reflected in the rate at which its localized Fourier transform decays, i.e., a function is smooth in a neighborhood of a point when the local Fourier transform decays rapidly. To put this formally, let $\xi_0 \in \Omega$  be a point in the domain and $\phi_{\xi_0}\in C_c^\infty(\Omega)$ be a \emph{cutoff function, }where $C_c^{\infty}(\Omega)$ denotes the space of infinitely differentiable functions with compact support in $\Omega$, and suppose that $\phi_{\xi_0}(\xi_0)\not = 0.$ Then the product $\phi_{\xi_0} x$ localizes $x$ to a neighborhood of $\xi_0$, since it agrees with $x$ near $\xi_0$ and vanishes outside the support of $\phi_{\xi_0}$. We say that $x$ is locally smooth at $\xi_0$ if its Fourier transform $\widehat{\phi_{\xi_0} x}(\omega)$ decay rapidly as $|\omega|\to\infty$, i.e., for any $N>0$, $|\widehat{\phi_{\xi_0} x}(\omega)| \le C_N (1+|\omega|)^{-N}$, as $|\omega| \to \infty$. Otherwise, if there are no such cutoff function, we refer to $\xi_0$ as a \emph{singularity} of $x$.

Microlocal analysis extends the notion of local smoothness by also
characterizing the directions in which singularities occur. These
directions will later help identify the optimal view-angles for
reconstruction. Let $\rho_0\in\mathbb{R}^2\setminus\{0\}$ denote a
direction (in the 2D plane) and let $V$ be an open wedge (i.e., an open conic
neighborhood) containing $\rho_0$. We say that $x$ is
\emph{microlocally smooth} at $(\xi_0,\rho_0)$ if there exist a cutoff
function $\phi_{\xi_0}\in C_c^\infty(\Omega)$ with
$\phi_{\xi_0}(\xi_0)\neq0$ and an open wedge $V$ containing $\rho_0$
such that, for every $N>0$, there exists a constant $C_N>0$ satisfying
\[
|\widehat{\phi_{\xi_0}x}(\omega)|
\le
C_N(1+|\omega|)^{-N},
\qquad \omega\in V.
\]
If no such cutoff function and open wedge exist, then $x$ is said to be
\emph{microlocally singular} at $(\xi_0,\rho_0)$. The collection of all
such position--direction pairs is called the \emph{wavefront set} of $x$
and is denoted by $\mathrm{WF}(x)$. The following lemma is an immediate consequence of the  theory
of wavefront sets (see, e.g., \cite{Quint-visibility}). We state it in
the setting of piecewise-constant images, which is sufficient
for our purposes.

\begin{lemma}\label{lem:Wf}
Let $\Omega\subset\mathbb{R}^2$, and let $x$ be a piecewise-constant
image whose discontinuities lie along a $C^1$ curve $\Gamma_{edge}$. Let
$\xi_0 =(\xi_1,\xi_2)\in\Gamma_{edge}$ be a point where the jump across $\Gamma_{edge}$ is nonzero.
Then
\[
\mathrm{WF}(x)\cap\bigl(\{\xi_0\}\times(\mathbb{R}^2\setminus\{0\})\bigr)
=
\{(\xi_0,\lambda n(\xi_0)):\lambda\in\mathbb{R}\setminus\{0\}\},
\]
where $n(\xi_0)$ is any nonzero vector normal to $\Gamma_{edge}$ at $\xi_0$.
In particular, each singular point has a unique wavefront direction,
up to nonzero scaling, which is orthogonal to the edge.
\end{lemma}

\begin{proof}
Since $\Gamma_{edge}$ is a $C^1$ curve, it has a unique tangent and normal direction at every point. Without loss of generality, we may assume $\xi_0=(0,0)$ and that $\Gamma_{edge}$ coincides locally with the line $\{\xi_2=0\}$. Since $x$ is piecewise-constant, it is locally equal to
\[
x(\xi)=
\begin{cases}
\alpha_1,&\xi_2>0,\\
\alpha_2,&\xi_2<0,
\end{cases}
\]
where $\alpha_1\neq \alpha_2$. This is a Heaviside-type jump across the line $\xi_2=0$. Its localized
Fourier transform fails to decay rapidly only in the normal direction
$(0,1)$, while it decays rapidly in all transverse directions.
Transforming back to the original coordinates, the unique wavefront
direction is the normal to $\Gamma_{edge}$ at $\xi_0$.
\end{proof}

Since
$(\xi_0,\rho)\in\mathrm{WF}(x)$ implies
$(\xi_0,-\rho)\in\mathrm{WF}(x)$, the wavefront direction is intrinsically
unoriented. Throughout the remainder of this work, we choose the unit outward normal to the corresponding inclusion boundary as a consistent representation of the wavefront direction.

Microlocal analysis characterizes how singularities of an attenuation field are mapped into its Radon transform. The Radon transform
$\mathcal R[x](\theta,s)$ is itself a function of the view-angle
$\theta \in [0,\pi)$ and detector coordinate $s$. Its singularities are described by
the wavefront set $\mathrm{WF}(\mathcal R[x](\theta,s))$. A classical result due to Quinto \cite{Quint-visibility} establishes a correspondence between
$\mathrm{WF}(x)$ and $\mathrm{WF}(\mathcal R[x](\theta,s))$.

\begin{theorem}[Adapted from Quinto~\cite{Quint-visibility}] \label{the:Q-visibility}
Let $x:\Omega\rightarrow\mathbb{R}$ and let
$\mathcal R[x](\theta,s)$ denote its Radon transform. If
$(\xi_0,\rho_0)\in\mathrm{WF}(x)$, then the corresponding
singularity is mapped to the wavefront set of the Radon transform at
the measurement
\[
(\theta,s),
\qquad
\boldsymbol{\theta}^\perp=\pm\frac{\rho_0}{\|\rho_0\|_2},
\qquad
s=\xi_0^\top\boldsymbol{\theta}^\perp.
\]
Equivalently, an image singularity is visible in the Radon transform
precisely at the view-angle that is orthogonal to the
wavefront direction.
\end{theorem}

In practice, the acquisition of measurements is often costly, time consuming, or physically limited. Therefore, it is important to determine a finite set of view-angles that maximizes the information content of the acquired measurements. In such cases, rather than considering the continuous Radon transform over all view-angles, we restrict our attention to a finite set of measurements $\boldsymbol{\theta}\in \mathbb [0,\pi)^N$ and use the discrete-angle Radon operator defined in \eqref{eq:finite-total-radon}.
The visibility result of Quinto's theorem immediately yields a corresponding visibility criterion for sparse-angle CT.

\begin{corollary} \label{cor:WForie}
Let $x:\Omega\rightarrow\mathbb{R}$ be a piecewise-constant image whose
discontinuities lie along $C^1$ curves, and let
$\Theta=\{\theta_i\}_{i=0}^{N_{\mathrm view}-1}\subset[0,\pi)$ denote the set of
measured view-angles. An edge point $\xi_0\in\Omega$ is visible in
the measured data if and only if its wavefront direction is orthogonal to one of the measured view directions, i.e.,
\[
\rho_0\perp\boldsymbol{\theta_i},     \qquad\text{for some component } \theta_i \text{ of }\Theta.
\]
\end{corollary}

In fact, the wavefront set not only identifies which singularities can be stably reconstructed from the available measurements, but also motivates the selection of view-angles from which singularities in the unknown image are visible in the sparse-angle setting.

\section{Bayesian optimal experimental design} \label{sec:BOED}
In Section \ref{sec:BayesForm}, we formulated the X-ray CT problem as a Bayesian inverse problem. Characterizing the resulting posterior distribution, for instance through its moments, provides a means of quantifying the spatially varying uncertainty that arises in the sparse-angle setting. In parallel, the microlocal analysis revisited in Section \ref{sec:microlocal-analysis} establishes a correspondence between singularities of the attenuation field and those of its sinogram, while Corollary \ref{cor:WForie} characterizes which singularities of the attenuation field are visible in the available sparse-angle sinogram data. Together, these Bayesian and microlocal perspectives naturally raise the following experimental design question: given the current posterior distribution associated with a sparse-angle X-ray CT problem, how can we optimally select the next view-angle so as to maximally reduce uncertainty in the recovery of singularities?

Bayesian optimal experimental design (BOED) is a well-established framework for selecting experimental design parameters, such as view-angles in X-ray CT, by optimizing an expected utility that quantifies the anticipated value of measurements under a given design \cite{Attia_2018, orozcoBOED,alexanderian_Review_InfD}. More precisely, an optimal design is typically defined as one that maximizes the expected utility or, equivalently, minimizes an expected loss or posterior uncertainty criterion. To put this formally, 
recall the image-space posterior distribution $\mu_{X \mid \boldsymbol Y_{\Theta}=\boldsymbol y_\Theta},$ introduced in \Cref{ssec:uq-posterior-sampling}, where $\Theta= \{\theta_i\}_{i=0}^{N_{\mathrm view}-1} \subset [0,\pi)$ denotes the current set of acquired view-angles.   Assuming that this measure has finite covariance, an A-optimality criterion quantifies the posterior uncertainty through the trace of the posterior covariance. Consequently, for a candidate view-angle $\theta$ and a prospective measurement $\boldsymbol Y_\theta$ the corresponding BOED problem is
\begin{equation} \label{general:A-opt}
\theta^{\ast}
\in
\operatorname*{arg\,min}_{\theta\in[0,\pi)}
\mathbb{E}_{\boldsymbol Y_{\theta} \mid \boldsymbol Y_{\Theta} = \boldsymbol y_\Theta,\theta}
\bigl[
\operatorname{tr}
\left(
\mathcal{C}_X^{\mathrm{post}}(\boldsymbol y_\Theta, \boldsymbol Y_\theta;\theta)
\right)
\bigr],
\end{equation}
where $\mathcal{C}_X^{\mathrm{post}}(\boldsymbol y_\Theta, \boldsymbol Y_\theta;\theta)$ denotes the posterior covariance of $X$ after conditioning on the current observation $\boldsymbol y_\Theta$ and the prospective measurement $\boldsymbol Y_\theta$.

While A-optimality seeks to minimize the overall posterior uncertainty in $X$, C-optimality, which can be viewed as a special case of goal-oriented A-optimality \cite{Attia_2018}, focuses on the uncertainty associated with a prescribed scalar quantity of interest (QoI). Let $G: L^2(\Omega) \to \mathbb{R}$ denote a linear functional acting on $X$, so that $G(X)$ represents the QoI. The corresponding experimental design is then chosen to minimize the expected posterior variance of $G(X)$, rather than that of $X$ itself. In particular, 
the posterior variance of the transformed quantity is $G\mathcal{C}^{\mathrm{post}}_XG^{*}$,  where
\(G^*:\mathbb R\to L^2(\Omega)\) denotes the adjoint of \(G\). The resulting design criterion takes the form
\begin{equation} \label{eq:c-optimality}
\theta^{\ast}
\in
\operatorname*{arg\,min}_{\theta\in[0,\pi)}
\mathbb{E}_{\boldsymbol Y_{\theta} \mid \boldsymbol Y_{\Theta}= \boldsymbol y_{\Theta},\theta}
\bigl[
\left(
G\mathcal{C}^{\mathrm{post}}_X(\boldsymbol y_{\Theta},\boldsymbol Y_{\theta};\theta) G^{*}
\right)
\bigr].
\end{equation}
Thus, the choice of $G$ allows the experimental design to target uncertainty
in a specific feature of the unknown rather than uncertainty in the entire attenuation field.

A central computational challenge in nonlinear BOED is that the posterior distribution, and consequently the posterior covariance, depends on both the experimental design and the as-yet-unobserved data. Evaluating the expected utility therefore generally requires integration over possible data realizations for each candidate design, resulting in a potentially high-dimensional nested integration and optimization problem. In this work, we formulate a microlocal C-optimal design criterion by introducing a QoI that localizes the posterior uncertainty around a spatial location of interest. 
The resulting criterion therefore seeks to reduce posterior uncertainty locally in a neighborhood of spatial location $\xi_0$. We show in Section \ref{sec:adaptive-angle-selection} that microlocal analysis can then be used to determine the view-angle that optimally reduces this localized uncertainty: the wavefront direction associated with $\xi_0$, together with the visibility relation in Corollary \ref{cor:WForie}, directly determines the subsequent view direction. Consequently, the proposed microlocal formulation replaces the outer numerical optimization over candidate view-angles with a geometrically determined acquisition rule. 

\section{Microlocally Informed Bayesian Optimal Experimental Design}   \label{sec:adaptive-angle-selection} 
In this section, we connect microlocal analysis with BOED in the image space by specializing the C-optimality criterion in \eqref{eq:c-optimality} to a scalar microlocal QoI associated with a spatial location and wavefront direction. We then exploit the structure of the star-shaped prior to construct efficient posterior surrogates for the B-MUSE algorithm.

Recall from \Cref{ssec:prior} that the attenuation field $X$  is modeled so that its realizations are piecewise-constant of the form of \eqref{eq:Xpiecewise}.  In this setting, the image singularities are precisely associated with the boundaries of the star-shaped inclusions. Moreover, this construction provides a direct interpretation of pointwise uncertainty in $X$ in terms of uncertainty
in the locations of the object boundaries. To see this, consider an object $D$ containing a single inclusion with fixed attenuation values $\alpha^+$ and $\alpha^-$. For any
$\xi\in\Omega$, the random attenuation field can be written as $X(\xi)=\alpha^-+(\alpha^+-\alpha^-)\boldsymbol{1}_{D}(\xi)$.
Conditioned on the measurements $Y_\Theta$, define
the posterior inclusion probability
\begin{equation}
p(\xi)
:=     
\mathbb{P}
\left(
\xi\in D
\mid
\boldsymbol Y_{\Theta}
\right).
\end{equation}
Since $\boldsymbol{1}_{D}(\xi)$ is a Bernoulli random variable with posterior
success probability $p(\xi)$, the pointwise posterior variance of $X$ is
\begin{equation}\
\operatorname{Var}
\left(
X(\xi)
\mid
\boldsymbol Y_{\Theta}
\right)
=
(\alpha^+-\alpha^-)^2
p(\xi)
\left(
1-p(\xi)
\right).
\label{eq:posterior-pointwise-variance}
\end{equation}

Consequently, the pointwise posterior variance is nonzero if and only if $0<p(\xi)<1$, that is, whenever posterior realizations disagree on whether $\xi$ belongs to the object $D$. Since the attenuation coefficients $\alpha^+$ and $\alpha^-$ are fixed and all uncertainty in $X$ is induced by the random geometry of $D$, this disagreement is entirely attributable to uncertainty in the location of the object boundary of $D$. Moreover, the posterior variance is maximized when $p(\xi)=1/2$, corresponding to spatial locations at which the posterior is maximally uncertain about whether $\xi$ lies inside or outside the object. Thus, in the piecewise-constant prior model considered here, the spatial posterior variance of $X$ provides a direct measure of uncertainty in the location of its singularities. The same argument extends to the multi-inclusion setting in \eqref{eq:Xpiecewise} when all inclusions have the same fixed attenuation coefficient $\alpha^+$.

The posterior distribution of $X$ induces a corresponding distribution on its wavefront set $\mathrm{WF}(X)$, describing posterior uncertainty in the locations and directions of its singularities. For the star-shaped prior considered here, the Mat\'ern covariance ensures that the inclusion boundaries are almost surely sufficiently regular \cite{stuart2010inverse}, and their wavefront directions are normal to the boundaries as characterized in Lemma~\ref{lem:Wf}. We denote the resulting \emph{posterior wavefront distribution} by
\begin{equation} \label{post-wf-dist}
\mu_{\mathrm{WF}(X)\mid \boldsymbol Y_{\Theta} = \boldsymbol y_{\Theta}}.
\end{equation}


Returning to the general piecewise-constant setting, the posterior distribution of $X$ and the posterior wavefront distribution characterize uncertainty in the locations and directions of its singularities, respectively. We first identify a location of maximal posterior uncertainty by
\begin{equation} \label{eq:posi-max-variance}
\xi_0
\in
\operatorname*{arg\,max}_{\xi\in\Omega}
\operatorname{Var}_{\mu_{X\mid \boldsymbol Y_\Theta=\boldsymbol y_\Theta}}
\left(
X(\xi)
\right).
\end{equation}

Now we connect the microlocal visibility criterion established in
\Cref{cor:WForie} with BOED. Specifically,
suppose that a point of maximal posterior uncertainty, $\xi_0$, as defined
in \eqref{eq:posi-max-variance}, coincides with the spatial location of a
true singularity, with associated wavefront direction $\rho_0$, so that $(\xi_0,\rho_0)\in \operatorname{WF}(x_{\mathrm{true}})$. Our aim is to show that, under this assumption, the view-angle satisfying
the microlocal visibility criterion, namely the angle orthogonal to $\rho_0$,
also yields the maximal reduction in posterior uncertainty associated with
the wavefront pair $(\xi_0,\rho_0)$.

To formulate this objective within the C-optimality framework of
\eqref{eq:c-optimality}, we introduce a scalar quantity of interest that is
localized both spatially near $\xi_0$ and directionally along $\rho_0$.
Let $\phi_{\xi_0}\in C_c^\infty(\Omega)$ be the cutoff function introduced
in Section~\ref{sec:microlocal-analysis}, and define the microlocal test function
\begin{equation}\label{eq:test_function}
\psi_{\xi_0,\rho_0}(\xi)
:=
\phi_{\xi_0}(\xi)
\exp
\left(
\boldsymbol{i}\rho_0^{\top}(\xi-\xi_0)
\right).
\end{equation}
We then define the scalar-valued linear functional
\begin{equation}\label{lin-C-func}
G_{\xi_0,\rho_0}(X)
:=
\left\langle
X,
\psi_{\xi_0,\rho_0}
\right\rangle_{L^2(\Omega)},
\end{equation}
which probes the local directional content of $X$ associated with the
wavefront pair $(\xi_0,\rho_0)$. Its posterior variance is given by
\[
G_{\xi_0,\rho_0}
\mathcal{C}^{\mathrm{post}}_X
G_{\xi_0,\rho_0}^{*}
=
\left\langle
\mathcal{C}^{\mathrm{post}}_X
\psi_{\xi_0,\rho_0},
\psi_{\xi_0,\rho_0}
\right\rangle_{L^2(\Omega)}.
\]

Substituting $G=G_{\xi_0,\rho_0}$ into the C-optimality criterion
\eqref{eq:c-optimality} yields the corresponding microlocal C-optimal
design problem
\begin{equation}\label{eq:microlocal-c-optimality}
\theta^{\ast}
\in
\operatorname*{arg\,min}_{\theta\in [0,\pi)}
\mathbb{E}_{\boldsymbol Y_\theta
\mid \boldsymbol Y_\Theta=\boldsymbol y_\Theta, \theta}
\left[
\left\langle
\mathcal{C}_X^{\mathrm{post}}
(\boldsymbol y_{\Theta},\boldsymbol Y_{\theta};\theta)
\psi_{\xi_0,\rho_0},
\psi_{\xi_0,\rho_0}
\right\rangle_{L^2(\Omega)}
\right].
\end{equation}
Thus, the design selects the next view-angle by minimizing the expected
posterior variance of the localized directional QoI
associated with $(\xi_0,\rho_0)$. The following result shows that, under
suitable local assumptions on the posterior covariance, this Bayesian
design criterion recovers the microlocal visibility condition.

\begin{theorem}[Microlocal C-optimality]
\label{thm:microlocal-c-optimality}
\label{thm:wavefront-optimality}
Let $(\xi_0,\rho_0)\in\operatorname{WF}(x_{\text{true}})$ be a wavefront pair of the true unknown image $x_{\text{true}}$, where
$\xi_0$ is a point of maximal posterior pointwise variance with its corresponding wavefront direction $\rho_0$, and let
$\psi_{\xi_0,\rho_0}$ and $G_{\xi_0,\rho_0}X$ be defined as in
\eqref{eq:test_function} and \eqref{lin-C-func}, respectively. Assume that:

\begin{enumerate}
    \item the current image-space posterior is locally well approximated by
    the Gaussian distribution
    \[
    X\mid\boldsymbol Y_\Theta=\boldsymbol y_\Theta
    \approx
    \mathcal N(m_X^{\mathrm{post}},\mathcal C_X^{\mathrm{post}});
    \]

    \item in a sufficiently small neighborhood of $\xi_0$, the posterior
    uncertainty is dominated by a single normal edge-displacement mode
    $v_{\xi_0,\rho_0}$, so that the localized posterior covariance admits
    the rank-one approximation
    \[
    \mathcal C_{\xi_0}^{\mathrm{post}}
    \approx
    \lambda_{\xi_0}
    v_{\xi_0,\rho_0}\otimes v_{\xi_0,\rho_0}, \qquad \lambda_{\xi_0} >0;
    \]

    \item the covariance contributions relevant to the localized QoI and the prospective measurement are dominated by this
    local mode; and

    \item the leading directional Fourier content of
    $v_{\xi_0,\rho_0}$ is concentrated along the wavefront direction
    $\rho_0$.
\end{enumerate}

Under these assumptions and the Gaussian noise model of \Cref{noise-Lklhood-post2}, the microlocal C-optimality criterion
reduces, to leading order, to
\begin{equation}
\operatorname*{arg\,max}_{\theta}
\Delta V_{\xi_0,\rho_0}(\theta)
\approx
\operatorname*{arg\,max}_{\theta}
\left\|
\mathcal R_\theta v_{\xi_0,\rho_0}
\right\|^2,
\end{equation}
where $\Delta V_{\xi_0,\rho_0}(\theta)$ denotes the reduction in goal-oriented posterior variance by incorporating $\theta$. Furthermore, the leading-order optimum satisfies $\theta^\star \perp \rho_0$.
Thus, the optimal view is tangent to the edge at $\xi_0$ and maximizes,
to leading order, the reduction in the posterior variance of the localized
directional quantity $G_{\xi_0,\rho_0}X$, consistently with
the microlocal visibility condition in Corollary~\ref{cor:WForie}.
\end{theorem}

\begin{proof}
The proof of \Cref{thm:microlocal-c-optimality} is given in
\Cref{app:variance-reduction} which shows that Gaussian conditioning reduces the microlocal
C-optimality criterion to maximizing the Radon energy of this local mode.
The Fourier-slice theorem then identifies the view that captures its
dominant directional Fourier content, yielding
$\theta^\star\perp\rho_0$, i.e., the view tangent to the edge at $\xi_0$.

For the star-shaped prior considered in
this work, conditions 2, 3, and 4 of the theorem are verified in
\Cref{app:star-shaped-local-mode}. In particular, a local perturbation of
a smooth star-shaped boundary is, to leading order, described by a single
normal-displacement mode, which gives rise to the local rank-one covariance
structure and directional Fourier concentration along $\rho_0$.
\end{proof}

In an inverse problem, the wavefront pair $(\xi_0,\rho_0)\in\mathrm{WF}(x_{\mathrm{true}})$ is generally unknown and must therefore be replaced by posterior-informed surrogates. The posterior distribution of $X$ provides a surrogate for the uncertain singularity location through \eqref{eq:posi-max-variance}, while the posterior wavefront distribution provides the corresponding directional information. Thus, the microlocal C-optimality result motivates selecting a view direction orthogonal to the posterior-informed wavefront direction associated with the region of greatest singularity uncertainty.

The practical construction of these surrogates can exploit the structure of the chosen prior. For the star-shaped prior considered in this work, uncertainty in singularity locations and directions can be represented directly through the inclusion index and boundary parameter $\varphi$. This allows us to replace computations over the full attenuation field by posterior quantities defined on the parameterized boundaries. In the following section, we use this structure to construct the numerical surrogates underlying the B-MUSE algorithm.

\subsection{B-MUSE: A Bayesian microlocally informed sequential experimental design algorithm} \label{sec:BMUSE}
In the previous section, we established the connection between microlocal analysis and BOED in terms of the posterior distribution of the attenuation field $X$ and its wavefront set. We now specialize this framework to a posterior constructed using the star-shaped prior. The boundary parameterization allows the posterior uncertainty in singularity locations and directions to be represented through the inclusion index $d$ and boundary angle $\varphi$, providing efficient posterior surrogates for the B-MUSE algorithm.

For numerical implementation, we discretize the angular interval $[0,2\pi)$ using $N_b$ equally spaced angles, $\{\varphi_j\}_{j=1}^{N_b}=\left\{2\pi (j-1)/N_b\right\}_{j=1}^{N_b}$. For each inclusion $d$, evaluating its star-shaped boundary parameterization at these angles maps every $\varphi_j$ to a spatial boundary point $\xi_d^j=(\xi_{1,d}^j,\xi_{2,d}^j)^\top\in\Omega$. Thus, the set $\{\xi_d^j\}_{j=1}^{N_b}$ provides a discrete spatial representation of the boundary of inclusion $d$ in the image domain. We also truncate the KL expansion \eqref{eq:KL} of $Z_d$ after $N_{\mathrm{KL}}$ terms, so that the prior, likelihood, and posterior can be formulated in finite-dimensional parameter and measurement spaces.


We now describe one iteration of the proposed adaptive Bayesian algorithm. At the beginning of each iteration $t$, we are provided with the current set of view-angles $\Theta_t$ and the corresponding, already acquired, sinogram. The corresponding posterior distribution is then explored using the NUTS sampler, generating $N_s$ posterior samples $\{(c_s,z_s)\}_{s=1}^{N_s}$, where each sample contains the parameters of all $N_{\mathrm{inc}}$ inclusions. 

Recall from the previous section that, under the piecewise-constant model, pointwise posterior uncertainty in $X$ is directly related to uncertainty in the location of the inclusion boundaries. For the star-shaped prior, we exploit this relation to construct a surrogate for the point of maximal uncertainty in \eqref{eq:post-var}. For each inclusion $d$ and boundary angle $\varphi_j$, every posterior sample $(c_s,z_s)$ determines a spatial boundary location $\xi_{s,d}^j:=\xi_d(\varphi_j;c_{s,d},z_{s,d})\in\Omega$. Thus, for fixed $(d,j)$, the collection $\{\xi_{s,d}^j\}_{s=1}^{N_s}$ represents the posterior variability in the spatial location of the boundary along the radial direction $\varphi_j$. We use this positional variability as a surrogate for the pointwise image-space uncertainty and seek the inclusion and boundary angle for which it is largest.

To quantify the positional variability of the boundary at each $(d,j)$, we define
\begin{equation}\label{eq:boundary-positional-variance}
V_{t,d}(\varphi_j):=\operatorname{tr}\!\left(\operatorname{Cov}\left(\{\xi_{s,d}^j\}_{s=1}^{N_s}\right)\right).
\end{equation}
Since $\xi_{s,d}^j\in\mathbb{R}^2$, this quantity is the trace of the $2\times2$ covariance matrix and measures the total spatial variability of the posterior boundary locations associated with $\varphi_j$.

The point of maximal boundary uncertainty is therefore identified by the inclusion and boundary angle satisfying
\begin{equation}\label{eq:alg-max-boundary-uncertainty}
(d_\star,j_\star)=\arg\max_{d,j}V_{t,d}(\varphi_j).
\end{equation}
An additional benefit of working with the boundary parameterization is that each posterior sample has a well-defined wavefront direction at every boundary angle $\varphi_j$. Hence, at the identified pair $(d_\star,j_\star)$, we estimate the expected posterior wavefront direction by averaging the wavefront vectors $\rho_{s,d_\star}^{j_\star}$ over the posterior samples and normalizing the resulting vector,
\begin{equation}\label{eq:alg-empirical-wavefront-direction}
\rho_\star=
\frac{\sum_{s=1}^{N_s}\rho_{s,d_\star}^{j_\star}}
{\left\|\sum_{s=1}^{N_s}\rho_{s,d_\star}^{j_\star}\right\|_2}.
\end{equation}

Motivated by the microlocal C-optimality result in \Cref{thm:microlocal-c-optimality}, we select the next view-angle so that its view direction is orthogonal to the estimated wavefront direction $\rho_\star$. Thus, the next B-MUSE projection angle is given by
\begin{equation}\label{eq:alg-next-view-angle}
\theta_{t+1}\quad \text{such that}\quad \boldsymbol{\theta}_{t+1}\perp\rho_\star.
\end{equation}
The next B-MUSE measurement is then acquired using the corresponding single view-angle measurement operator $\mathcal R_{\theta_{t+1}}$ defined in \eqref{eq:single-angle-R}. The new projection is added to the available measurements, the posterior is updated, and the procedure is repeated to determine the next view-angle. This procedure is summarized in \Cref{alg:BMUSE} with stopping criteria introduced in the following section.

\begin{algorithm}[!t]
\caption{B-MUSE: Bayesian microlocally informed sequential experimental
design for uncertainty reduction}
\label{alg:BMUSE}

\begin{algorithmic}[1]
\Require Initial view-angle set $\Theta_0$, initial measurements
$y_{\Theta_0}$, number of posterior samples $N_s$, number of boundary
points $N_b$, $N_{res}$ , $N_{sat}$ 
\Ensure Adaptively selected view-angle set $\Theta_t$

\State Set $t \gets 0$

\While{true}

    \State Construct the current forward operator
    $\mathcal{R}_{\Theta_t}$ according to \eqref{eq:finite-total-radon}.

    \State Construct the likelihood using $\boldsymbol y_{\Theta_t}$ and the
    corresponding posterior distribution according to \eqref{eq:likelihood} and \eqref{eq:posterior} respectively.

    \State Draw $N_s$ posterior samples
    $\{(c_s,z_s)\}_{s=1}^{N_s}$ using NUTS.

    \For{$s=1,\ldots,N_s$}
        \State Compute the attenuation-field realization
        $x_s=F(c_s,z_s)$ using \eqref{eq:F-Sigmoid}.
        \For{$d=1,\ldots,N_{\mathrm{inc}}$}
            \For{$j=1,\ldots,N_b$}
                \State Evaluate the boundary point
                $\xi_{s,d}^{\,j}$ at the angular node $\varphi_j$.
                \State Compute the corresponding unit outward
                wavefront direction $\rho_{s,d}^{\,j}$.
            \EndFor
        \EndFor
    \EndFor

    \State Compute the pixel-wise posterior variance according to
    \eqref{eq:posterior-pointwise-variance}.

    \State Compute the posterior boundary positional variance
    $V_{t,d}(\varphi_j)$ according to \eqref{eq:boundary-positional-variance}.

    \State Evaluate the stopping criteria
    \eqref{stop1} and \eqref{eq:stop2-A}.

    \If{either stopping criterion is activated}
        \State \textbf{break}
    \EndIf

    \State Identify the inclusion and boundary point with maximum
    positional uncertainty according to \eqref{eq:alg-max-boundary-uncertainty}, obtaining
    $(d^\star,j^\star)$.

    \State Compute the corresponding posterior mean boundary location
    $\xi^\star$.

    \State Estimate the associated posterior mean wavefront direction
    $\rho^\star$.

    \State Select the next view-angle $\theta_{t+1}$ such that
    \( \boldsymbol {\theta}_{t+1}\perp\rho^\star,\)
    according to Theorem \ref{thm:microlocal-c-optimality}.

    \State Acquire the new measurement $y_{\theta_{t+1}}$.

    \State Update
    \(    \Theta_{t+1}
       =\Theta_t\cup\{\theta_{t+1}\},
    \qquad
    y_{\Theta_{t+1}}
       =y_{\Theta_t}\cup y_{\theta_{t+1}}.  \)

    \State Set $t\gets t+1$.

\EndWhile

\State \Return $\Theta_t$

\end{algorithmic}
\end{algorithm}

\subsection{Stopping criteria}
Our proposed \Cref{alg:BMUSE} selects the most informative view direction at the end of each iteration. However, an adaptive algorithm whose aim is to optimize number of view-angles, should determine not only which measurement to acquire next, but also when the current acquired measurements are sufficient. The amount of information required depends on both the object and the measurement noise level. 
We therefore design an adaptive stopping criterion  according to the current posterior uncertainty and the assumed statistical noise model. In this formulation, we use only quantities that are available.    First, we use a resolution  criterion and terminate the acquisition once the posterior positional uncertainty of the boundary falls below the spatial resolution of one grid. Second, for cases in which this resolution level is not reached, we use a saturation criterion and terminate the acquisition only when new view-angle produce small changes in reducing relative positional boundary uncertainty.

Recall that $V_{t,d}(\varphi_j)$ in \eqref{eq:boundary-positional-variance} quantifies the posterior positional uncertainty of the boundary of inclusion $d$ at the angular parameter $\varphi_j$ and iteration $t$. Once this uncertainty falls below the spatial resolution of the image discretization, further reductions cannot be resolved on the computational grid. Therefore, for the posterior conditioned on the measurements $\boldsymbol y_{\Theta_t}$, we stop when
\begin{equation}
    \label{stop1}
    \mathcal{V}_t:=\max_{\substack{1\leq d\leq N_{\mathrm{inc}}\\
                    1\leq j\leq N_b}}
    V_{t,d}(\varphi_j)
    <
    \left(\frac{L}{N_{\mathrm{pix}}-1}\right)^2 .
\end{equation}
Here, $L/(N_{\mathrm{pix}}-1)$ is the spatial grid spacing for a domain of side length $L$. Since $\mathcal{V}_t$ may fluctuate between iterations, we activate this stopping criterion only when \eqref{stop1} is satisfied for $N_{res}$ consecutive iterations.
This criterion is motivated by the fact that requiring the posterior positional uncertainty to decrease substantially below the pixel scale would provide little additional information. Moreover, one can scale this criterion based on the needed resolution in their studies. 

However, under high measurement noise, the maximum positional uncertainty may not reach below the resolution grid. In such cases, considering \eqref{stop1} as the only stopping criterion may lead to unnecessary measurements without any reduction.
We therefore consider the second stopping rule as a metric for this reduction. 

At iteration $t$ the number of available 
measurements is $N_{\mathrm{detect}}|\Theta_t|$. Since we are assuming a Gaussian noise model, for the normalized squared noise we have  
$
\frac{\|\varepsilon_{\Theta_t}\|_2^2}{\sigma^2}
\sim \chi^2_{N_{\mathrm{detect}}|\Theta_t|}.
$
Hence, 
we have
\[
\mathbb{E}[\frac{\|\varepsilon_{\Theta_t}\|_2^2}{(N_{\mathrm{detect}}|\Theta_t|_2)\sigma^2}]=1,
\quad
\operatorname{std}[\frac{\|\varepsilon_{\Theta_t}\|^2}{(N_{\mathrm{detect}}|\Theta_t|)\sigma^2}]=
\sqrt{\frac{2}{N_{\mathrm{detect}}|\Theta_t|}}.
\]
This quantity characterizes the fluctuation scale of the normalized noise under the assumed Gaussian noise model \cite{Iglesias_2021}. Motivated by this, we use the same scale as a reference for the relative change in the maximum positional uncertainty. Specifically, we define
\begin{equation}
    \label{eq:stop2-A}
    \frac{ \left| \sqrt{V_t}-\sqrt{V_{t-1}} \right|}{\sqrt{V_{t-1}}}  < \sqrt{\frac{2}{N_{\mathrm{detect}}|\Theta_t|}}.
\end{equation}
This condition is satisfied, when the relative change in the maximum positional  uncertainty between two consecutive iterations is smaller than the scale motivated by the normalized noise. We interpret that as a sign that  additional view is unlikely to produce a huge uncertainty reduction. However, we avoid terminating the acquisition at the first time; we activate this criterion only when  \eqref{eq:stop2-A} is satisfied for $N_{sat}$ consecutive iterations. 

Overall, the acquisition terminates either when the posterior positional uncertainty
reaches the spatial resolution of the computational grid for $N_{res}$ consecutive iterations, or when its relative change remains below the noise-calibrated statistical scale for $N_{sat}$  consecutive iterations.

\section{Numerical experiments} \label{sec:Numerical-r}
In this section, we investigate the performance of the B-MUSE algorithm. To this end, we conduct four numerical experiments. In the first experiment, we consider a disk as the ground truth inclusion. The disk geometry provides a simple setting for illustrating our method, since its boundary is simple and its wavefront directions are radial. The ground truth in the second experiment is a realization of the star-shaped prior introduced in \Cref{ssec:prior}. Compared to a disk, this inclusion has a more complex boundary and wavefront directions that vary along the boundary.  Therefore, it allows us to examine the effectiveness of our method in reducing uncertainty. In the third study, we benchmark our method against other related experimental design methods to compare the resulting sparse-view acquisition designs and reconstruction accuracy under the same measurement budget. In the fourth experiment, our goal is to evaluate our algorithm in the presence of both model mismatch and multiple inclusions.

\subsection{Setup} \label{ssec:numerical-setup}
We perform all experiments on the image domain \( \Omega=[-1,1]^2, \) discretized on an $N_{\mathrm{pix}}\times N_{\mathrm{pix}}$ Cartesian grid with $N_{\mathrm{pix}}=128$. 
In \eqref{eq:finite-total-radon}, we defined the total measurement operator \(\mathcal R_{\Theta}\) for a finite number of view-angles $N_{\mathrm{view}} $ and detector locations $N_{\mathrm{detect}}$. Its input is a continuous attenuation field, and each measurement is defined by a line integral. For numerical experiments, we use $N_{\mathrm{detect}}=128$  and evaluate these line integrals on the discretized grid using $N_{\mathrm{quad}}=128$ quadrature points per ray.
We set the attenuation coefficients to \(\alpha^-=0\) and \(\alpha^+_d=1\) for \(d=1,\ldots,N_{inc},\) where $N_{inc}$ denotes the number of objects in the piecewise-constant representation \eqref{eq:Xpiecewise}. 
We use the star-shaped prior introduced in \Cref{ssec:prior} throughout the numerical experiments, with experiment-specific restrictions described in the corresponding subsections. For experiments involving the star-shaped priors, we truncate the KL-expansion of $Z$  after \(N_{\mathrm{KL}}=256\) terms in \eqref{eq:F-Sigmoid}. For these experiments, we assign the Mat\'ern parameters \(\kappa=10\) and \(\alpha=4\) to the covariance of $Z$ to produce sufficiently smooth boundary realizations. Following the boundary discretization introduced in \Cref{sec:BMUSE}, we use equally spaced \(N_b=256\) angular nodes, providing a fine angular spacing on the boundary relative to the radius range and the image $\Omega$. 
%
We begin each experiment with the single horizontal view-angle $\theta_0=0.$ Therefore, the initial  view-angle set is \( \Theta_0= \{\theta_0\}=\{0\}.\) The corresponding sinogram contains \( N_\mathrm{sino}= N_{\mathrm{detect}}= 128\) measurements. At each iteration, \Cref{alg:BMUSE} selects one view-angle.  After $t$ iterations, the angle set becomes $\Theta_t=\{\theta_0,\theta_1,\ldots,\theta_{t}\},$ with \(N_{\mathrm{view}}=t+1.\)  The corresponding sinogram then contains \( N_\mathrm{sino}= 128 \times (t+1)\). 
Moreover, we consider $N_{res}= 3$ and $N_{sat}=4$ to activate the stopping criteria.  
To assess the effect of measurement noise on our method, we consider three relative noise levels of $0.1\%,$ $5\%$ and $10\%,$ and generate observations according to \eqref{eq:additive-noise-model}. Let \(\eta\in\{0.001,0.05,0.10, \}\) denote the relative noise level. For each noise level,
we set 
 \( \sigma=\sigma_\eta :=  \eta\, \|\mathcal{R}_{\Theta_\mathrm{full}}(x_{\mathrm{true}}) \|_\infty, \) where $\Theta_\mathrm{full} =\left\{
\frac{j\pi}{180}
\right\}_{j=0}^{179}
\subset[0,\pi).$  Here, we use the same standard deviation for all acquired view-angles. Moreover, the noise level is not dependent to the selection of the initial view-angle. Hence, the noise model in \eqref{noise-model} becomes \( \boldsymbol   E_{\theta_i} \sim \mathcal{N} \left(0,\sigma_\eta^2 I_{N_{\mathrm{detect}}}  \right), \ i=0,\ldots,N_{\mathrm{view}}-1. \) 
For a direct comparison, we use the same 
realization of standardized noise vectors and only change the noise scale  $\sigma_\eta.$

For all experiments, we explore the posterior distribution using NUTS with one Markov chain, 800 warm-up steps, and 2000 posterior samples. To assess sampling efficiency of the Markov chain, we compute the mean effective sample size (ESS) from the posterior samples \cite{mcbook} . We also report the cumulative runtime of each experiment up to the stopping iteration. 
All computations are performed on the CPU of a MacBook Pro with an Apple M4 Pro chip and 24 GB of unified memory, using Python 3.9.6, PyTorch 2.7.1, and Pyro 1.9.1 for posterior sampling. 

\subsection{Optimal design for a disk inclusion} \label{ssec:disk-experiment}
In this experiment, we  consider an image containing one disk inclusion with center $c_{\mathrm{true}}=(0.17,-0.10)$ and radius $b_{\mathrm{true}}=0.25$ as the ground truth, i.e., $N_{inc}=1.$ 
The corresponding piecewise-constant true attenuation field is
\begin{equation}
    x_{\mathrm{true}}(\xi)
    =
    \begin{cases}
        1,
        & \|\xi-c_{\mathrm{true}}\|_2\leq b_{\mathrm{true}}\\
        0,
        & \|\xi-c_{\mathrm{true}}\|_2>b_{\mathrm{true}}
    \end{cases}, \qquad \xi \in \Omega.
    \label{eq:disk-attenuation}
\end{equation}  
Its singularities lie on the disk boundary, which is a \(C^\infty\) curve. By Lemma \ref{lem:Wf}, its wavefront set is supported on the disk boundary, with orientation perpendicular to the boundary at that point. Since we adapted the outward orientation in this work, these directions are aligned with the radial direction from the center of the disk. 

For reconstructing this inclusion, we consider a disk prior as a special case of the star-shaped prior introduced in \Cref{ssec:prior}. We restrict it to the constant mode of KL-expansion in \eqref{eq:KL}.  We set the latent random variable $Z$ to be constant for all angles $\varphi \in [0,2\pi),$ i.e., $Z(\varphi)=A,\ \varphi \in [0,2\pi)$ where $A\sim\mathcal{N}\big(\log(0.1),1\big).$ Therefore,  this prior generates only disk-shaped inclusions.

The following Proposition \ref{Disk2angles} explains that two nonparallel view-angles are sufficient to determine a disk. The first projection determines its radius and one directional component of its center. A second nonparallel projection then determines the remaining center component. Through this simple geometry of disk, we can examine whether the selected angle is maximally reducing the uncertainty in the boundary of the disk.  



\begin{proposition}\label{Disk2angles}
In a piecewise-constant attenuation field, a disk-shaped ground truth inclusion is uniquely determined by two different nonparallel view-angles.
\end{proposition}
\begin{proof}
Let the attenuation field $x$ be as \eqref{eq:disk-attenuation} a piecewise-constant function with the inclusion center $c$ and radius $b.$ 
For one view direction $\boldsymbol\theta_0=\begin{pmatrix}
        \cos\theta_0&\sin\theta_0
    \end{pmatrix}^T$, evaluating  the Radon
transform gives
\[   \mathcal{R}[x_{c,b}](\boldsymbol\theta_0,s)
=
\begin{cases}
2\sqrt{
b^2-\left(s-c^\top\boldsymbol\theta_0^\perp\right)^2
},
&
\left|s-c^\top\boldsymbol\theta_0^\perp\right|\leq b,
\\[1mm]
0,
&
\left|s-c^\top\boldsymbol\theta_0^\perp\right|>b.
\end{cases} \] This single angle measurement is nonzero only for detector locations satisfying
\(     c^\top\boldsymbol\theta_0^\perp-b    \leq s \leq  c^\top\boldsymbol\theta_0^\perp+b.\)
Let \(s_0^{-}\) and \(s_0^{+}\) denote the two endpoints of
this interval. Then \(  b
    =
    \frac{s_0{+}-s_0^{-}}{2},
    \
    c^\top\boldsymbol\theta_0^\perp
    =
    \frac{s_0^{+}+s_0^{-}}{2}. \)
Thus, one projection determines the radius and one component of the center. By another view-angle $\boldsymbol\theta_1=\begin{pmatrix}
        \cos\theta_1&\sin\theta_1
    \end{pmatrix}^T$, we arrive at the  equation $   \begin{pmatrix}
    \boldsymbol\theta_0^\perp &\boldsymbol\theta_1^\perp
\end{pmatrix} ^T c= \begin{pmatrix}
        \dfrac{s_0^{+}+s_{0}^{-}}{2}&
        \dfrac{s_{1}^{+}+s_{1}^{-}}{2}
    \end{pmatrix}^T.$ For determining $c$, the matrix $\begin{pmatrix}
   \boldsymbol \theta_0^\perp &\boldsymbol\theta_1^\perp
\end{pmatrix} ^T$ must be invertible. Hence by nonparallel $\boldsymbol\theta_0$ and $\boldsymbol\theta_1$ a disk is fully determined. Moreover, the determinant \( 
   \big| \det
    \begin{pmatrix}
        (\boldsymbol\theta_0^\perp)&
        (\boldsymbol\theta_1^\perp)
    \end{pmatrix}^T \big|  \) is maximized when 
        \(  |\theta_1-\theta_0 |= \frac{\pi}{2}. \) 
    Therefore the best reconstruction is obtained when the two view-angles are orthogonal. 
\end{proof}

Proposition \ref{Disk2angles} provides a simple benchmark to check whether B-MUSE behaves consistently with theory. Figure \ref{fig:Disk-iter1,2} illustrates the first two iterations of the disk experiment for the three different noise levels. It shows the posterior reconstruction, posterior uncertainty, and corresponding view-angles together with the new view-angle proposed by the B-MUSE \Cref{alg:BMUSE}.

\begin{figure}[t!]
    \centering    \includegraphics[width=\linewidth]{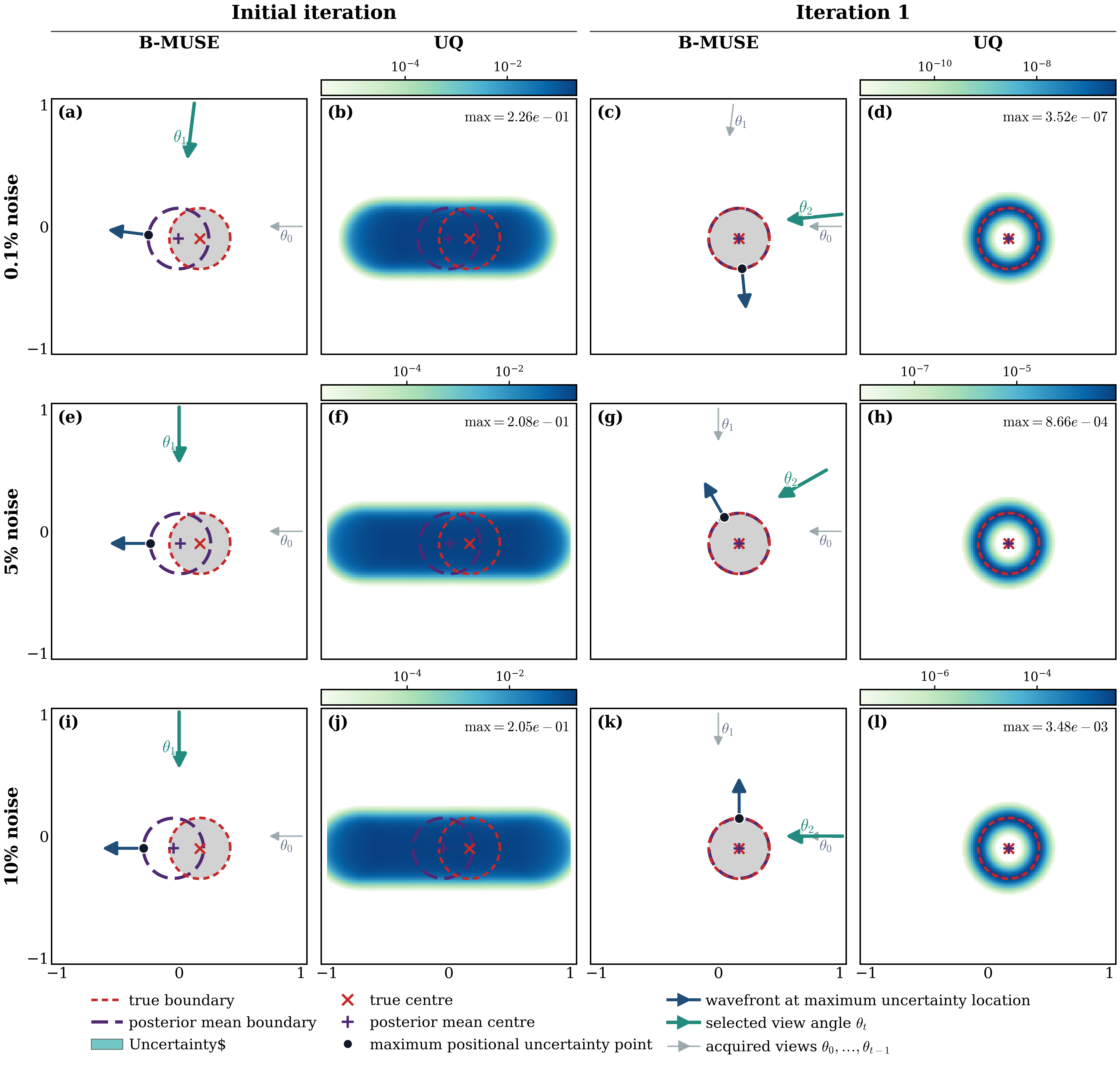}
    \caption{First two iterations of the  disk experiment for the
    \(0.1\%\), \(5\%\), and \(10\%\) noise levels. The B-MUSE columns show the true and posterior mean boundaries, the true and posterior mean centers, the point of maximum positional variance, its associated wavefront direction, the acquired view-angles, and the
    selected view-angle. The UQ columns show the pixel-wise posterior variance and its maximum value. Here, the 
    the initial iteration uses \(\theta_0=0\), whereas iteration \(1\)
    includes \(\theta_1\) .}
    \label{fig:Disk-iter1,2}
\end{figure}

In the initial iteration of this experiment, the reconstruction is based on one horizontal view-angle $\theta_0=0$. As predicted by Proposition \ref{Disk2angles}, this view-angle determines disk radius and one component of the center, while the vertical component of the center remains unconstrained. Consequently, the posterior uncertainty spreads horizontally, which is consistent with Corollary \ref{the:Q-visibility}. Therefore, posterior mean boundary is displaced from the true boundary. 
Then based on the expected wavefront direction on the most uncertain location, B-MUSE selects  $\theta_1$, which is approximately orthogonal to the $\theta_0$ in all three noise levels. This agrees with both Proposition \ref{Disk2angles} and Theorem \ref{thm:microlocal-c-optimality}.
After assimilating the view-angle $\theta_1$, the posterior mean center moves very close to the true center, and the uncertainty concentrates as a narrow band around the boundary. This behavior is consistent with Proposition \ref{Disk2angles}. In all three noise levels, this behavior is analogous, although the magnitude of the remaining posterior uncertainty increases with the higher noise levels. 
To quantify the effect of each proposed view-angle, we track the error in the estimated disk center, the error in the estimated radius, and the maximum positional variance, as seen from \Cref{tab:disk-errors}. At iteration $t$, the posterior is conditioned on  $\boldsymbol y_{\Theta_{t}}$. We define the  radius and center relative errors as 
\(   e_B:= \|
 \mathbb E[ B \mid \boldsymbol y_{\Theta_{t}}]-b_{\mathrm{true}}  \|_2/
\|b_{\mathrm{true}
}\|_2 \) 
 and \( e_C:= \| \mathbb E[C \mid \boldsymbol y_{\Theta_{t}}]-c_{\mathrm{true}} \|_2 / {\|c_{\mathrm{true}}\|_2} 
,  \) respectively. 

\begin{table}[H]
\caption{Relative errors and maximum  positional uncertainty for the
three noise regimes. The variance reduction $\Delta V$ is computed from the
initial iteration to iteration 1.}
\label{tab:disk-errors}
\centering
\renewcommand{\arraystretch}{0.70}
\begin{tabular}{@{}llcccc@{}}
\hline
Noise &
Stage &
$e_{B}$ &
$e_{C}$ &
$\max_j V_t(\varphi_j)$ &
$\Delta V$ (\%)
\\
\hline

$0.1\%$ &
Initial iteration &
0.0051 &
0.8863 &
$1.1125\times10^{-1}$ &
\multirow{2}{*}{$>99.9999$}
\\
&
Iteration 1 &
0.0059 &
0.0036 &
$3.6817\times10^{-9}$ &
\\
\hline

$5\%$ &
Initial iteration &
0.0089 &
0.8037 &
$1.8793\times10^{-1}$ &
\multirow{2}{*}{99.9950}
\\
&
Iteration 1 &
0.0088 &
0.0028 &
$9.3526\times10^{-6}$ &
\\
\hline

$10\%$ &
Initial iteration &
0.0123 &
1.0984 &
$2.0132\times10^{-1}$ &
\multirow{2}{*}{99.981}
\\
&
Iteration 1 &
0.0138 &
0.0047 &
$3.8400\times10^{-5}$ &
\\
\hline

\end{tabular}
\end{table}
\Cref{tab:disk-errors} shows that B-MUSE removes most of the uncertainty with the first proposed angle. This is the result of a significant reduction in center error and radius error. These results are based on posterior samples at iteration $t=1$, for which the mean ESS values are $1817$, $2368$, and $1849$ for the $0.1\%$, $5\%$, and $10\%$ noise levels respectively.  The maximum runtime for this experiment was approximately $3$ minutes.


\subsection{Optimal design for a prior sample}
The ground truth inclusion in this experiment is a sample of the prior distribution introduced in \Cref{ssec:prior}. Its center is \(c_{\mathrm{true}}=(0.2,0)^{T} ,\) and its radial function is one realization of truncated $Z$ after $N_{KL}=256$ term with Mat\'ern covarience.   The corresponding true attenuation field \(x_{\mathrm{true}}\) is then obtained by mapping \(F_H(C,Z)\) in \eqref{eq:F-Heaviside}, which is piecewise-constant field. We also used $\beta= 50$ in \eqref{eq:F-Sigmoid} when we explore the posterior. Since this boundary is a \(C^1\) curve, by Lemma \ref{lem:Wf}, the singularities of \(x_{\mathrm{true}}\) lie on the inclusion boundary and their wavefront directions are normal to the boundary. 
Unlike the disk prior, for which the wavefront direction is radial at every boundary point, the wavefront direction of a star-shaped inclusion is not necessarily aligned with the radial direction. Reconstructing the inclusion in this example is more difficult than in the disk experiment because instead of a baseline radius, the radial function must be also inferred.  This experiment therefore enables us to evaluate whether B-MUSE method can identify spatially varying boundary, its uncertainty, and select view-angles to mitigate this uncertainty.

\begin{figure}[t!]
    \centering
\includegraphics[width=1.\linewidth]{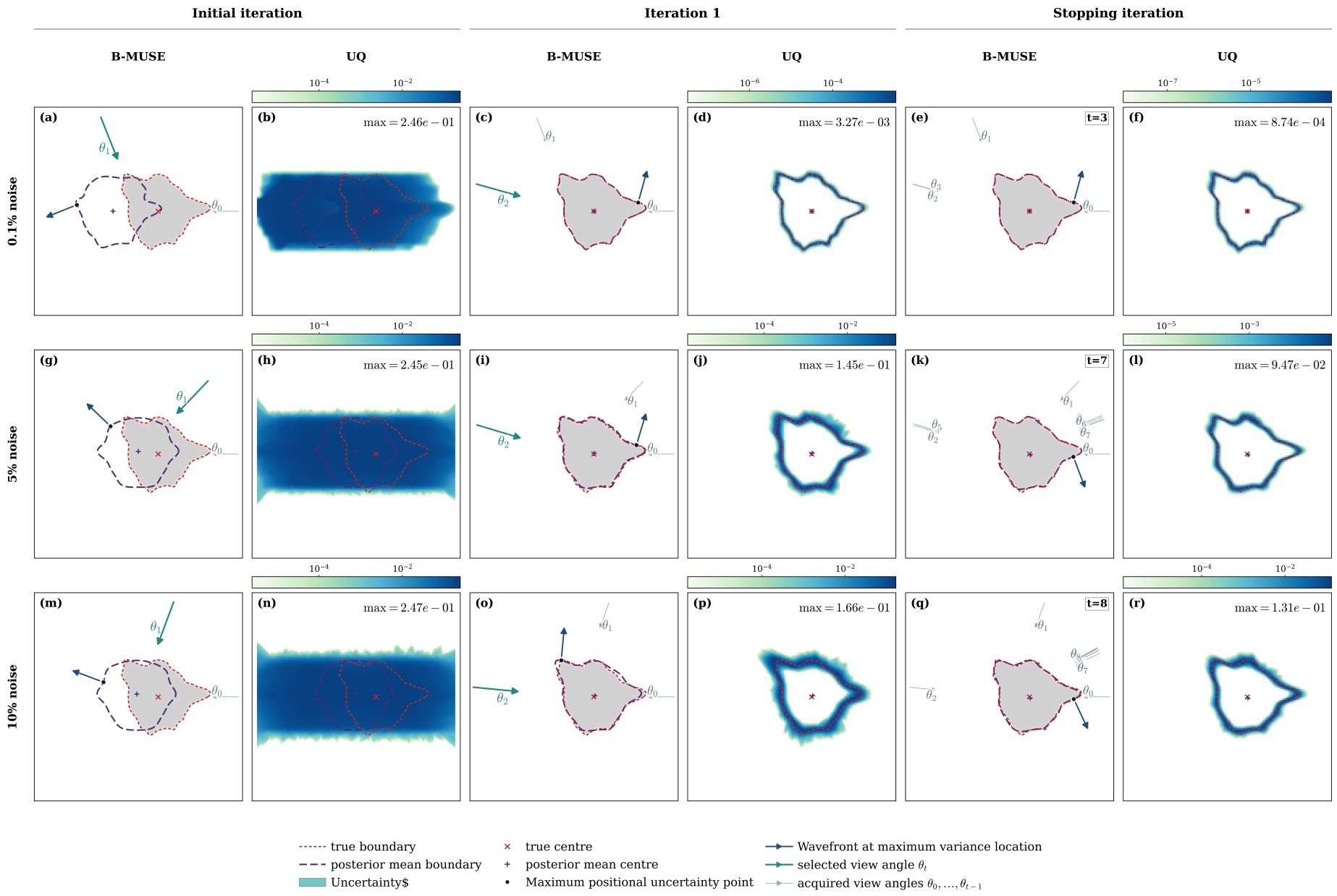}
    \caption{First two iterations and the stopping iteration in the prior sample experiment for  the
    \(0.1\%\), \(5\%\), and \(10\%\) noise levels. The B-MUSE columns show the true and posterior mean boundaries, the true and posterior mean centers, the point of maximum positional uncertainty, its associated wavefront direction, the acquired view-angles, and the
    selected view-angle. The UQ columns show the pixel-wise posterior variance and its maximum value. The initial iteration uses only $\theta_0=0$, iteration $1$ includes $\theta_1$, and the stopping iteration $t$ uses the view-angle set
\(
\Theta_t\).}
\label{fig:prior-sample}
\end{figure}
Figure~\ref{fig:prior-sample} shows the result of B-MUSE in the First two iterations and the stopping iteration for the \(0.1\%\), \(5\%\),
and \(10\%\) noise levels.  At the initial iteration, measurements are acquired based on $\theta_0$, thus the uncertainty spreads over a horizontal region. B-MUSE proposes the next view-angle $\theta_1$ according to \Cref{alg:BMUSE} to optimally tackle this uncertainty. After adding $\theta_1,$ the reconstruction significantly improves, and the posterior mean boundary moves closer to the true boundary, while the associated uncertainty region narrows. 
 The maximum positional variance decreases by $99.97\%$, $99.41\%$, and $98.96\%$ for the $0.1\%$,  $5\%$, and $10\%$ noise levels, respectively, relative to its initial value. The corresponding reductions in the maximum pixel-wise posterior variance are $98.67\%$, $40.79\%$, and $32.80\%$. Thus, even at higher noise levels, boundary becomes more localized, while some pixel-wise uncertainty remains near it. 
 
\begin{figure}[t!]
    \centering
    \includegraphics[width=\linewidth]{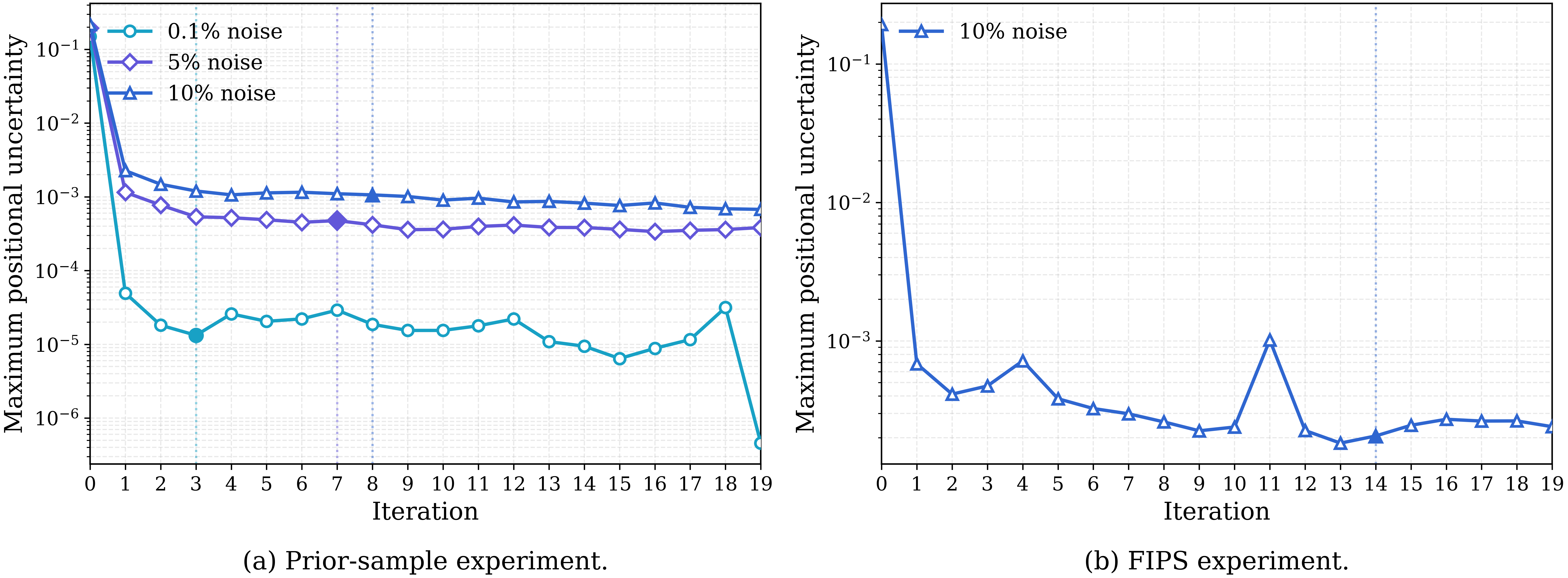}
    \caption{Maximum positional variance over the iterations for the
    Prior-sample and FIPS experiments.}
    \label{fig:max-var-prior-fips}
\end{figure}

 The effect of assimilating the proposed view-angle is strongest at the lowest noise level, where both pixel-wise and positional uncertainties are 
reduced.  At higher noise levels, the same 
behavior is observed, although a wider uncertainty band remains around the reconstructed boundary. During the next iterations, 
proposed view-angles refine the reconstruction, and the stopping criteria are reached at iterations $t=3$, $7$, and $8$ for the \(0.1\%\), \(5\%\),
and \(10\%\) noise levels, respectively.  This means that, for $0.1\%$, the maximum positional uncertainty was smaller than the pixel resolution after the initial iteration and remained below this value for three next iterations, hence, \eqref{stop1} was activated.
For \(5\%\), and \(10\%\) noise levels, the positional uncertainty did not reach the pixel resolution. The second stop criterion \eqref{eq:stop2-A} is activated, meaning that the relative change in maximum positional uncertainty for 4 consecutive iterations remaind below the chosen value.
Posterior samples at these stopping iterations had  mean ESS value $1678$, $1973$, and $4063$ for the $0.1\%$, $5\%$, and $10\%$ noise levels, respectively.
 The maximum cumulative runtime through this iteration was approximately $4.4$ hours.
Figure \ref{fig:max-var-prior-fips}(a) shows the variation of the maximum positional uncertainty. The maximum positional uncertainty in the initial iteration is approximately the same across all noise levels. 
After acquiring $\boldsymbol y_{\Theta_1}$, the largest reduction occurs, but the extent of this reduction depends on the noise level. In the next iterations, positional
uncertainty reduces with a slower rate comparing to the initial reduction.
In particular, after removing the highest uncertainty in the first iteration, a plateau is reached with some fluctuations. Thus, the termination indicates that the stopping criterion in \eqref{stop1} and  \eqref{eq:stop2-A} has been satisfied, rather than that further uncertainty reduction is impossible. Although continuing the acquisition may reduce this uncertainty, it requires additional radiation exposure. A trade-off therefore arises between acquiring further information about the boundary and limiting
the radiation dose while accepting some remaining uncertainty. We suggest future work to adapt these criteria to the accuracy
requirements and radiation constraints of a specific application.

The overall trend of the maximum positional uncertainty is decreasing, however it is not monotone. After updating the posterior with a new noisy measurement, the uncertainty at a boundary location may increase, since the new measurements can make alternative boundaries plausible. Moreover,  only on average over all possible measurements the posterior variance is guaranteed to decrease, not necessarily for each  measurement realization. Another reason for nonmonoton behavior is that we estimate the positional variance  from a limited number of NUTS samples, which may introduce small fluctuations.  B-MUSE uses  posterior surrogates, this may also contribute to some fluctuations.

\subsection{Comparison with Alternative Experimental Design Methods}
To put the performance of B-MUSE into perspective, we compare it with two other optimal experimental design approaches: projection-based approximation of conditional expectation (PACE) \cite{PACE} and neural optimal design of experiments (NODE) \cite{Node}. Both approaches replace computationally demanding components of conventional optimal experimental design with neural approximations and permit gradient-based optimization of continuous experimental-design parameters. In the X-ray CT setting considered here, the design variables are the view-angles. To ensure a controlled comparison, NODE and PACE are trained using samples generated from the same prior distribution employed by B-MUSE, rather than using an external training dataset.

PACE and NODE provide two learning-based approaches to experimental design. PACE is a Bayesian A-optimal design method that uses the law of total variance and the projection characterization of conditional expectation to replace posterior computations with a neural regression approximation of the conditional mean, yielding a differentiable objective for optimizing the view-angles. In contrast, NODE jointly optimizes a neural reconstruction map and the acquisition angles using a reconstruction loss; in our sparse-view CT experiments, a U-Net is trained together with three view-angles, producing a fixed three-view acquisition protocol that is subsequently applied to the test object.

For both NODE and PACE, training samples are generated on the fly from the same prior distribution used by B-MUSE. At each training step, prior samples are propagated through the differentiable Radon transform at the current design angles to generate noisy measurements. To ensure a consistent noise level across all methods (including B-MUSE) and experiments, we use additive Gaussian noise with a fixed standard deviation for each acquired view, independent of the signal magnitude, rather than the signal-dependent noise model introduced earlier. Thus, no external dataset is used, and both the networks and design variables are trained for 150 epochs.


Standard NODE and PACE are non-adaptive: all three view-angles are optimized with respect to the prior predictive distribution before observing the test object and are then fixed for acquisition. In contrast, B-MUSE updates the design sequentially using measurements of the specific object. To assess this benefit separately, we also consider adaptive NODE (aNODE), initialized with the same prescribed view $\theta_0=0$ as B-MUSE. After acquiring this measurement, aNODE jointly trains the reconstruction network and optimizes the next angle $\theta_1$ conditioned on the available observation; after acquiring $\theta_1$, the two measurements are fixed and the final angle $\theta_2$ is optimized. Thus, the aNODE design ${\theta_0,\theta_1,\theta_2}$ is sequential and object-dependent, with $\theta_1$ and $\theta_2$ informed by previously acquired measurements.

We compare the four design strategies on the disk inclusion and the fixed realization of the star-shaped prior. PACE, NODE, and aNODE reconstruct the attenuation image, from which the inclusion boundary is extracted as the $0.5$-level set. In contrast, B-MUSE operates directly on the low-dimensional parameters of the star-shaped inclusion, with the image and boundary obtained through the prior push-forward map. The comparison therefore favors B-MUSE in terms of representation, since it directly targets the QoI while the neural methods must first reconstruct the full image.

\begin{figure}[t!]
    \centering

    \begin{subfigure}{\linewidth}
        \centering
        \includegraphics[width=\linewidth]{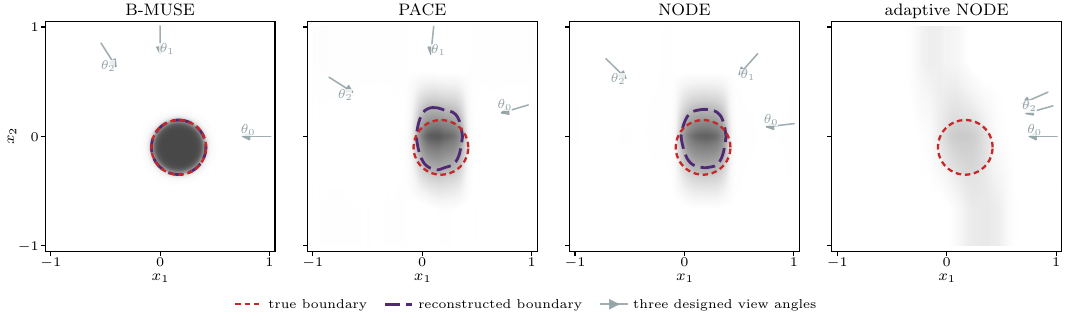}
        \caption{Comparison of B-MUSE, PACE, NODE, and adaptive NODE for the disk inclusion.}
        \label{fig:compare_NODE_PACE_disk}
    \end{subfigure}

    \vspace{0.5em}

    \begin{subfigure}{\linewidth}
        \centering
        \includegraphics[width=\linewidth]{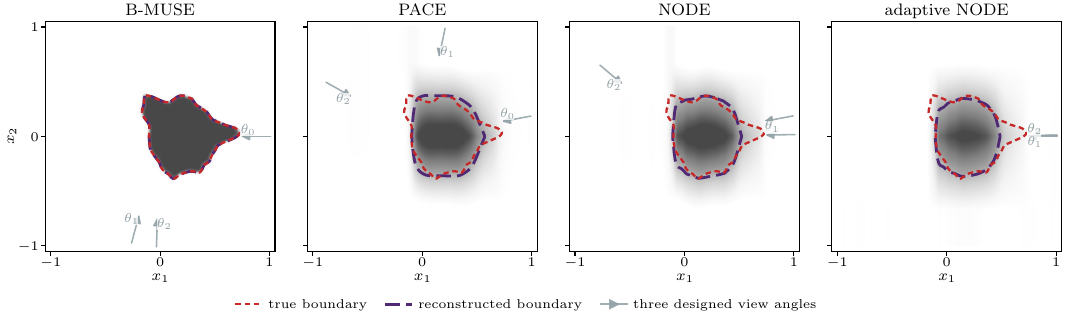}
        \caption{Comparison of B-MUSE, PACE, NODE, and adaptive NODE for the star-shaped prior sample.}
        \label{fig:compare_NODE_PACE_star}
    \end{subfigure}

    \caption{Comparison of the optimal experimental designs and corresponding reconstructions for the disk and star-shaped inclusion problems at $1\%$ noise.}
    \label{fig:compare_NODE_PACE}
\end{figure}

The disk phantom results are shown in \Cref{fig:compare_NODE_PACE_disk}. With only three views, B-MUSE recovers the inclusion location and boundary more accurately than the neural approaches. PACE and NODE produce comparable reconstructions but select different acquisition geometries, with PACE giving more regular angular spacing and NODE a more asymmetric design. In contrast, aNODE selects views with limited angular diversity and consequently provides the least accurate boundary reconstruction.

The star-shaped phantom results in \Cref{fig:compare_NODE_PACE_star} show similar behavior. B-MUSE captures the principal boundary variations, whereas PACE and NODE recover the location and overall shape but produce smoother boundaries that miss finer radial features. This may partly reflect their pixel-wise mean-squared-error training objective, which gives relatively little weight to localized boundary errors, while B-MUSE represents uncertainty directly in the boundary parameterization.

The non-adaptive designs show similar angular patterns to the disk experiment: PACE selects comparatively regular view spacing, while NODE produces a more asymmetric configuration. Although aNODE improves substantially on the star-shaped phantom, its selected angles remain strongly clustered, suggesting that the reconstruction improvement may partly reflect the learned U-Net prior rather than complementary information from the additional views. Separating these effects would require an ablation with fixed or repeated view-angles.

Finally, reconstruction from only three views represents an exceptionally challenging sparse-view CT setting and is not intended as a general assessment of NODE or PACE. Rather, it probes an extreme information-limited regime in which B-MUSE can exploit its low-dimensional geometric prior and sequential uncertainty-informed design. For the boundary-reconstruction task considered here, these features provide a substantial advantage over full-image neural reconstruction and design.


\subsection{Optimal design for a multi-inclusion phantom}
In this section, we evaluate B-MUSE on a multi-inclusion phantom, which is not generated by our assumed prior in \ref{ssec:prior}. The ground truth phantom in this experiment is taken from the Helsinki Tomography Data Challenge 2022 dataset \cite{fipsHTC2022}. It contains four inclusions, each with a sufficiently regular boundary. The ground truth is represented as a piecewise-constant function of the form given in  \eqref{eq:Xpiecewise}. Python codes for this experiments are accessible through \cite{giahi-sabour-bmuse}

For the reconstruction, we use $N_{inc}=4$ , base radius $b_0=0.18$, radial scaling $s_B= 0.05$ and the sigmoid parameter $\beta=50$ in \eqref{eq:F-Sigmoid}. We restricted the inclusion centers to $[-0.75,0.75]^2$, and use a pairwise repulsion to prevent inclusions from collapsing into just one same inclusion given by
\begin{equation*}
\begin{aligned}
&\mathcal{R}_{\mathrm{rep}}({\{c_d\}_{d=1}^{N_{\text{inc}}}})
=
\sum_{i=1}^{N_{\mathrm{inc}}-1}
\sum_{j=i+1}^{N_{\mathrm{inc}}}
\left[
\frac{1}{\beta_{\text{rep}}}
\log\left(
1+\exp\left[
\beta_{\text{rep}}\left(d_{\min}-d_{ij}\right)
\right]
\right)
\right]^2,
\\
&d_{ij}
=
\|c_i-c_j\|_2+\varepsilon,
\quad
\varepsilon=10^{-8}, \quad d_{\text{min}} = 0.15, \quad
\beta_{\text{rep}} = 30.
\end{aligned}
\end{equation*}
Since we already examined the affect of noise, in this experiment we  study the highest noise level considered in other experiments, $10\%$.
\begin{figure}[t] \centering
\includegraphics[width=0.98\linewidth]{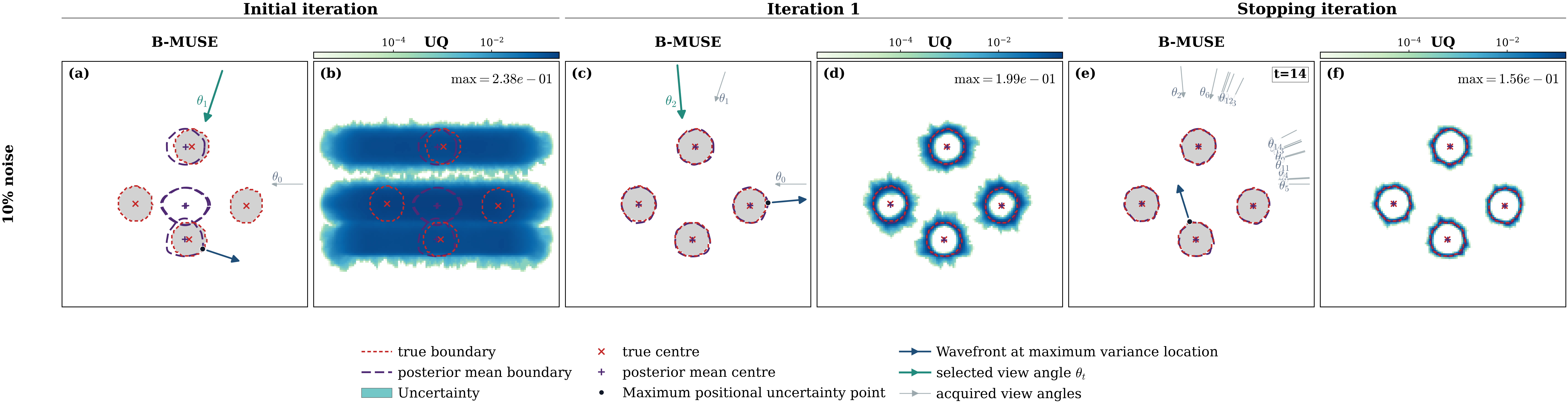}
    \caption{First two iterations and the stopping iteration of a multi-inclusion example from an external dataset for the \(10\%\) noise level. The B-MUSE columns show the true and posterior mean boundaries, the true and posterior mean centers, the point of maximum positional uncertainty, its associated wavefront direction, the acquired view-angles, and the
    selected view-angle. The UQ columns show the pixel-wise posterior variance and its maximum value. The initial iteration uses only $\theta_0=0$, iteration $1$ includes $\theta_1$, and the stopping iteration $t$ uses the view-angle set
\(\Theta_t\). Each UQ plot uses an independent logarithmic color scale.}
    \label{fig:fips-sample}
\end{figure}


Figure \ref{fig:fips-sample} shows the initial iteration, the first adaptive iteration, and the stopping iteration for the $10\%$ noise level.
In the initial iteration,  the posterior uncertainty spreads horizontally along   the four inclusions, where the initial view-angle $\theta_0=0$ is tangent to some boundary point of inclusions. Posterior mean boundaries therefore are displaced from the true boundaries.
Then B-MUSE identifies the boundary point with the largest positional variance over all four inclusions and chooses the next view-angle according to \Cref{alg:BMUSE}. Consequently,  the uncertainty in the next iteration with assimilating $\theta_1$ is concentrated around boundaries. However, some inclusions remain displaced. As the acquisition continues, B-MUSE sequentially selects next
view-angles using the positional uncertainty across the inclusion boundaries. At the stopping iterations $t = 14$ based on criteria \eqref{stop1}, the posterior mean
boundaries closely follow the four true boundaries, although a small
pixel-wise uncertainty remains.


Figure \ref{fig:max-var-prior-fips}(b) shows the variation of the maximum positional uncertainty. At the initial iteration, the maximum positional uncertainty is \(1.9297\times10^{-1}\).
After acquiring measurements $\boldsymbol y_{\Theta_1}$, this value
decreases to \(6.8390\times10^{-4}\), corresponding to relative
reduction about \(99.65\%\), meaning that 
 the first proposed angle removes the dominant portion of the posterior uncertainty. 
The maximum pixel-wise posterior variance also decreases by approximately \(16.39\%\) from the initial iteration to the first iteration.
In the next iterations, the maximum positional variance decreases but in a smaller extent, with few marginal exceptions. The stopping criterion is activated by the resolution
criterion, after the positional uncertainty remains below the 
pixel scale for the three consecutive iterations.
At the stopping iteration \(t=14\), the maximum
positional uncertainty is \(2.0604\times10^{-4}\), corresponding to approximately \(99.89\%\) reduction  relative to the initial
iteration. 
The posterior samples at this iteration had a mean ESS of $4058$, and the cumulative runtime  was approximately $6.2$ hours.

\section{Conclusion}\label{sec:conclusion}
In this work, we introduced B-MUSE, a goal-oriented Bayesian experimental design method for sequential view-angle selection in sparse-angle X-ray CT. Here, we considered the inclusion boundaries as the QoIs and used the posterior moments to reconstruct them and quantify their uncertainty. We defined a positional uncertainty value along the boundary using samples drawn by NUTS. This new uncertainty allows us to identify the boundary location that remains most uncertain under the current measurements. Our proposed method B-MUSE, at each iteration, selects this location and estimates its corresponding posterior mean wavefront direction; then it selects a new view-angle that makes the uncertain singularity visible to the measurement.

We formulated the connection between visibility and uncertainty reduction through a goal-oriented microlocal C-optimality criterion. We showed that the optimal reduction in the uncertainty of the posterior is when the next view-angle is orthogonal to the wavefront direction at the location of highest uncertainty. In this sense, B-MUSE is able not only to characterize the visible singularities according to current measurements, but also to determine which measurement should be acquired next. 

Our conducted experiments showed the behavior of B-MUSE under different geometries and noise levels. 
For the disk example, the selected measurements agreed with the analytical result.  For more general star-shaped boundaries, B-MUSE selects the view-angles according to posterior positional uncertainty. Moreover, in the out-of-prior multi-inclusion experiment, B-MUSE remained effective and obtained a significant reduction in posterior uncertainty within the first two iterations alone. 

Our results show that the available measurements should not be treated equally. Instead, selecting their directionality according to the unresolved information in the QoIs can provide an accurate reconstruction with considerably lower uncertainty. Stopping criteria can also be adapted to the level of uncertainty that suits the specific application purposes. This can potentially reduce radiation exposure in X-ray CT. In applications such as tumor contouring, where the boundary itself is the QoIs, B-MUSE can be a useful tool, since it can continue the acquisition only until the uncertainty in the contour reaches an acceptable level. Moreover, B-MUSE is model-based and does not require a training dataset.  Although the model covers piecewise-constant attenuation fields, with a prior capable of encoding the wavefront set, it can be extended to piecewise-smooth attenuation fields as well.

As a direction for future research, B-MUSE can be extended to other piecewise-constant prior models, such as level-set priors. More broadly, the underlying microlocal principles are not restricted to piecewise-constant images. Since piecewise-smooth images also exhibit singularities along interfaces, a substantial part of the B-MUSE framework may be extended to piecewise-smooth prior models. This is particularly relevant for goal-oriented X-ray CT applications in which accurate reconstruction of inclusion boundaries is more important than recovering fine-scale variations within the background or foreground. In such settings, B-MUSE could potentially provide a sparse-angle acquisition design that prioritizes accurate boundary reconstruction while allowing smoother estimation away from the interfaces.

\appendix

\section{Variance reduction under an additional projection}
\label{app:variance-reduction}
In this appendix, we derive the reduction in the posterior variance of the localized quantity of interest introduced in \Cref{sec:adaptive-angle-selection} when an additional view-angle is acquired. The derivation is carried out under a Gaussian approximation of the current posterior distribution of the attenuation field, which provides an explicit expression for the dependence of the posterior variance reduction on the candidate view direction.

Recall that the current image-space posterior is the push-forward
\begin{equation}
\mu_{X\mid \boldsymbol Y_\Theta=\boldsymbol y_\Theta}
=
F_{\#}\mu_{(C,Z)\mid \boldsymbol Y_\Theta=\boldsymbol y_\Theta},
\end{equation}
and is generally non-Gaussian due to the nonlinear parameter-to-image map $F$. For the following local variance-reduction analysis, we approximate it by a Gaussian distribution with the same posterior mean and covariance,
\begin{equation}
X\mid \boldsymbol Y_\Theta=\boldsymbol y_\Theta
\approx
\mathcal N\left(
m_X^{\mathrm{post}},
\mathcal C_X^{\mathrm{post}}
\right),
\label{eq:app-gaussian-posterior}
\end{equation}
where $m_X^{\mathrm{post}}$ and $\mathcal C_X^{\mathrm{post}}$ can be estimated from posterior samples. This approximation is motivated locally by posterior concentration together with the differentiability of $F$, which permits a first-order linearization around a posterior reference point.


We focus on a point $\xi_0$ of maximal posterior pointwise variance,
\begin{equation}
\xi_0
\in
\operatorname*{arg\,max}_{\xi\in\Omega}
\operatorname{Var}
\left(
X(\xi)\mid\boldsymbol Y_\Theta=\boldsymbol y_\Theta
\right).
\end{equation}
Assume that $\xi_0$ lies on a smooth edge of the true attenuation field, and let $\rho_0\in\mathbb R^2\setminus\{0\}$ denote its wavefront direction, i.e., the direction normal to the edge at $\xi_0$. We show that, under the local approximations introduced below, the view tangent to this edge maximizes the leading-order reduction of the posterior uncertainty associated with $(\xi_0,\rho_0)$.

To quantify this uncertainty, we use the localized directional test function introduced in equation~\eqref{eq:test_function},
\begin{equation}
\psi_{\xi_0,\rho_0}(\xi)
:=
\phi_{\xi_0}(\xi)
\exp\left(
\boldsymbol{i}\rho_0^\top(\xi-\xi_0)
\right),
\label{eq:app-microlocal-test}
\end{equation}
where $\phi_{\xi_0}\in C_c^\infty(\Omega)$ localizes the test to a neighborhood of $\xi_0$, while the oscillatory factor probes the direction $\rho_0$. We define the corresponding scalar quantity of interest by
\begin{equation}
G_{\xi_0,\rho_0}X
:=
\left\langle
X,\psi_{\xi_0,\rho_0}
\right\rangle_{L^2(\Omega)}.
\label{eq:app-qoi}
\end{equation}
Thus, $G_{\xi_0,\rho_0}X$ measures the image content localized near $\xi_0$ and oriented in the wavefront direction $\rho_0$. Under the Gaussian approximation in equation~\eqref{eq:app-gaussian-posterior}, its current posterior variance is
\begin{equation}
V_{\Theta}(\xi_0,\rho_0)
:=
\operatorname{Var}
\left(
G_{\xi_0,\rho_0}X
\mid
\boldsymbol Y_\Theta=\boldsymbol y_\Theta
\right)
=
\left\langle
\mathcal C_X^{\mathrm{post}}\psi_{\xi_0,\rho_0},
\psi_{\xi_0,\rho_0}
\right\rangle_{L^2(\Omega)}.
\label{eq:app-current-qoi-variance}
\end{equation}

We now consider the acquisition of one additional projection at a candidate angle $\theta\in[0,\pi)$. Recall that the single view-angle measurement operator \eqref{eq:single-angle-R} is $\mathcal R_\theta:L^2(\Omega)\rightarrow\mathbb R^{N_{\mathrm{detect}}}$. The corresponding prospective measurement is the random vector
\begin{equation}
\boldsymbol Y_\theta=\mathcal R_\theta X+\boldsymbol E_\theta,
\label{eq:app-new-measurement}
\end{equation}
where the noise follows the model introduced in \Cref{noise-Lklhood-post2}, i.e., $\boldsymbol E_\theta\sim\mathcal N\left(0,\sigma^2I_{N_{\mathrm{detect}}}\right)$, and $\boldsymbol E_\theta$ is independent of $X$ and of the measurement noise associated with the previously acquired projections.

We seek the candidate angle $\theta$ that maximally reduces the posterior uncertainty in the localized quantity of interest $G_{\xi_0,\rho_0}X$. Under the Gaussian approximation \eqref{eq:app-gaussian-posterior} and the linear measurement model \eqref{eq:app-new-measurement}, define the posterior predictive covariance
\begin{equation}
\mathcal S_\theta
:=
\mathcal R_\theta
\mathcal C_X^{\mathrm{post}}
\mathcal R_\theta^*
+
\sigma^2 I_{N_{\mathrm{detect}}}.
\label{eq:app-predictive-covariance}
\end{equation}
Since $\sigma^2>0$, $\mathcal S_\theta$ is positive definite and therefore invertible. The standard Gaussian conditioning formula, equivalently the Schur-complement formula for the conditional covariance, gives the posterior covariance after acquiring the additional projection,
\begin{equation}
\mathcal C_X^{\mathrm{post},\theta}
=
\mathcal C_X^{\mathrm{post}}
-
\mathcal C_X^{\mathrm{post}}
\mathcal R_\theta^*
\mathcal S_\theta^{-1}
\mathcal R_\theta
\mathcal C_X^{\mathrm{post}}.
\label{eq:app-covariance-update}
\end{equation}
Within this linear-Gaussian approximation, $\mathcal C_X^{\mathrm{post},\theta}$ depends on the candidate angle $\theta$ but not on the prospective data realization $\boldsymbol y_\theta$. Thus, candidate angles can be compared directly through the resulting reduction in the posterior variance of $G_{\xi_0,\rho_0}X$.

The posterior variance of the localized quantity of interest after acquiring the additional projection at angle $\theta$ is
\begin{equation}
V_{\Theta\cup\{\theta\}}(\xi_0,\rho_0)
=
\left\langle
\mathcal C_X^{\mathrm{post},\theta}
\psi_{\xi_0,\rho_0},
\psi_{\xi_0,\rho_0}
\right\rangle_{L^2(\Omega)}.
\label{eq:app-updated-qoi-variance}
\end{equation}
We define the corresponding variance reduction by
\begin{equation}
\Delta V_{\xi_0,\rho_0}(\theta)
:=
V_{\Theta}(\xi_0,\rho_0)
-
V_{\Theta\cup\{\theta\}}(\xi_0,\rho_0).
\label{eq:app-variance-reduction-definition}
\end{equation}
Substituting the covariance update \eqref{eq:app-covariance-update} and using the self-adjointness of $\mathcal C_X^{\mathrm{post}}$ together with the adjoint relation for $\mathcal R_\theta$ gives
\begin{equation}
\Delta V_{\xi_0,\rho_0}(\theta)
=
\left\langle
\mathcal S_\theta^{-1}
\mathcal R_\theta
\mathcal C_X^{\mathrm{post}}
\psi_{\xi_0,\rho_0},
\mathcal R_\theta
\mathcal C_X^{\mathrm{post}}
\psi_{\xi_0,\rho_0}
\right\rangle_{\mathbb R^{N_{\mathrm{detect}}}}.
\label{eq:app-variance-reduction-inner-product}
\end{equation}

Since $\mathcal S_\theta$ is symmetric positive definite, it admits a unique symmetric positive-definite inverse square root $\mathcal S_\theta^{-1/2}$. Therefore, the variance reduction can be expressed as
\begin{equation}
\Delta V_{\xi_0,\rho_0}(\theta)
=
\left\|
\mathcal S_\theta^{-1/2}
\mathcal R_\theta
\mathcal C_X^{\mathrm{post}}
\psi_{\xi_0,\rho_0}
\right\|_{\mathbb R^{N_{\mathrm{detect}}}}^2.
\label{eq:app-variance-reduction}
\end{equation}

Since $V_\Theta(\xi_0,\rho_0)$ is independent of the candidate angle $\theta$, minimizing the updated posterior variance is equivalent to maximizing its reduction. Hence, the microlocal C-optimal design problem can be written as
\begin{equation}
\theta^\ast
\in
\operatorname*{arg\,max}_{\theta\in[0,\pi)}
\Delta V_{\xi_0,\rho_0}(\theta)
=
\operatorname*{arg\,max}_{\theta\in[0,\pi)}
\left\|
\mathcal S_\theta^{-1/2}
\mathcal R_\theta
\mathcal C_X^{\mathrm{post}}
\psi_{\xi_0,\rho_0}
\right\|_{\mathbb R^{N_{\mathrm{detect}}}}^2.
\label{eq:app-variance-reduction-design}
\end{equation}

To characterize the dominant posterior uncertainty associated with the selected edge, we first isolate the part of the posterior covariance localized near $\xi_0$. This localization will allow us to identify the leading local mode of uncertainty and, subsequently, to approximate the local posterior covariance by a rank-one operator.

Let $\phi_{\xi_0}\in C_c^\infty(\Omega)$ denote the spatial cutoff used in the definition of $\psi_{\xi_0,\rho_0}$, and let
\[
M_{\phi_{\xi_0}}u:=\phi_{\xi_0}u.
\]
We define the posterior covariance localized to the neighborhood of $\xi_0$ by
\begin{equation}
\mathcal C_{\xi_0}^{\mathrm{post}}
:=
M_{\phi_{\xi_0}}
\mathcal C_X^{\mathrm{post}}
M_{\phi_{\xi_0}}^*.
\label{eq:app-localized-covariance}
\end{equation}
Since $\phi_{\xi_0}$ is real-valued, $M_{\phi_{\xi_0}}^*=M_{\phi_{\xi_0}}$. Thus, $\mathcal C_{\xi_0}^{\mathrm{post}}$ describes the posterior covariance of image perturbations restricted to the neighborhood selected by $\phi_{\xi_0}$.

We next characterize the dominant posterior fluctuation represented by
$\mathcal C_{\xi_0}^{\mathrm{post}}$. Rather than specializing immediately
to the star-shaped parameterization, we formulate the local property needed
for the variance-reduction argument directly in image space.

We assume that, in a sufficiently small neighborhood of the smooth edge
through $\xi_0$, the posterior uncertainty is dominated to first order by
a single local boundary-displacement mode. More precisely, after
localization by $\phi_{\xi_0}$,
\begin{equation} \label{eq:app-local-one-mode} 
M_{\phi_{\xi_0}}
\left(
X-m_X^{\mathrm{post}}
\right)
=
\widetilde A_{\xi_0}
v_{\xi_0,\rho_0}
+
r_\varepsilon,
\end{equation}
where $\widetilde A_{\xi_0}$ is a centered scalar random amplitude,
$v_{\xi_0,\rho_0}\in L^2(\Omega)$ is the image-space mode associated
with a local displacement of the edge in its normal direction $\rho_0$,
and $r_\varepsilon$ is the local approximation remainder, with
$\varepsilon$ denoting the size of the localization neighborhood. We assume that
\begin{equation}
\mathbb E
\left[
\|r_\varepsilon\|_{L^2(\Omega)}^2
\mid
\boldsymbol Y_\Theta=\boldsymbol y_\Theta
\right]
\longrightarrow 0,
\qquad
\varepsilon\longrightarrow 0.
\label{eq:app-local-one-mode-remainder}
\end{equation}
We further assume that the leading directional Fourier content of
$v_{\xi_0,\rho_0}$ is concentrated along the normal direction $\rho_0$,
as is characteristic of a localized smooth edge displacement.

This local one-mode property is the only feature of the geometric prior
needed in the argument below. In Appendix~\ref{app:star-shaped-local-mode},
we show how it arises from the star-shaped parameterization used in this
work.

The local one-mode assumption \eqref{eq:app-local-one-mode} immediately
provides a rank-one approximation of the localized posterior covariance.
Indeed, neglecting the remainder $r_\varepsilon$ to leading order gives
\begin{equation}
\mathcal C_{\xi_0}^{\mathrm{post}}
\approx
\lambda_{\xi_0}
v_{\xi_0,\rho_0}
\otimes
v_{\xi_0,\rho_0},
\qquad
\lambda_{\xi_0}
:=
\operatorname{Var}
\left(
\widetilde A_{\xi_0}
\mid
\boldsymbol Y_\Theta=\boldsymbol y_\Theta
\right).
\label{eq:app-local-rank-one-covariance}
\end{equation}
Thus, to leading order, the localized posterior uncertainty is concentrated
along the single image-space mode $v_{\xi_0,\rho_0}$.



Assuming that the covariance contributions relevant to the localized
quantity of interest and the prospective measurement are dominated by this
local mode, we have
\begin{equation}
\mathcal C_X^{\mathrm{post}}\psi_{\xi_0,\rho_0}
\approx
\mathcal C_{\xi_0}^{\mathrm{post}}\psi_{\xi_0,\rho_0},
\end{equation}
and
\begin{equation}
\mathcal S_\theta
\approx
\lambda_{\xi_0}
\left(\mathcal R_\theta v_{\xi_0,\rho_0}\right)
\left(\mathcal R_\theta v_{\xi_0,\rho_0}\right)^*
+
\sigma^2I.
\label{eq:app-local-predictive-rank-one}
\end{equation}
Substituting \eqref{eq:app-local-rank-one-covariance} into the
variance-reduction criterion gives
\begin{equation}
\Delta V_{\xi_0,\rho_0}(\theta)
\approx
\lambda_{\xi_0}^2
\left|
\left\langle
\psi_{\xi_0,\rho_0},
v_{\xi_0,\rho_0}
\right\rangle_{L^2(\Omega)}
\right|^2
\left\langle
\mathcal S_\theta^{-1}
\mathcal R_\theta v_{\xi_0,\rho_0},
\mathcal R_\theta v_{\xi_0,\rho_0}
\right\rangle.
\label{eq:app-local-variance-reduction-rank-one}
\end{equation}

Let $q_\theta:=\mathcal R_\theta v_{\xi_0,\rho_0}$. Then
$\mathcal S_\theta\approx\sigma^2I+\lambda_{\xi_0}q_\theta q_\theta^*$,
and the Sherman--Morrison formula \cite{sherman1950adjustment} gives
\begin{equation}
\left\langle
\mathcal S_\theta^{-1}q_\theta,q_\theta
\right\rangle
=
\frac{\|q_\theta\|^2}
{\sigma^2+\lambda_{\xi_0}\|q_\theta\|^2}.
\end{equation}
Consequently,
\begin{equation}
\Delta V_{\xi_0,\rho_0}(\theta)
\approx
\lambda_{\xi_0}^2
\left|
\left\langle
\psi_{\xi_0,\rho_0},
v_{\xi_0,\rho_0}
\right\rangle_{L^2(\Omega)}
\right|^2
\frac{
\|\mathcal R_\theta v_{\xi_0,\rho_0}\|^2
}{
\sigma^2+
\lambda_{\xi_0}\|\mathcal R_\theta v_{\xi_0,\rho_0}\|^2
}.
\label{eq:app-reduced-variance-reduction}
\end{equation}
Since the final factor is strictly increasing in
$\|\mathcal R_\theta v_{\xi_0,\rho_0}\|^2$, it follows that
\begin{equation}
\operatorname*{arg\,max}_{\theta}
\Delta V_{\xi_0,\rho_0}(\theta)
\approx
\operatorname*{arg\,max}_{\theta}
\left\|
\mathcal R_\theta v_{\xi_0,\rho_0}
\right\|^2.
\label{eq:app-radon-energy-objective}
\end{equation}

It remains to identify the view-angle maximizing the Radon energy in
\eqref{eq:app-radon-energy-objective}. Let
\begin{equation}
n_\theta
=
(\cos\theta,\sin\theta)^\top,
\qquad
n_\theta^\perp
=
(-\sin\theta,\cos\theta)^\top.
\end{equation}
By the Fourier-slice theorem and Plancherel's identity,
\begin{equation}
\left\|
\mathcal R_\theta v_{\xi_0,\rho_0}
\right\|^2
=
\int_{\mathbb R}
\left|
\widehat v_{\xi_0,\rho_0}
\left(
s n_\theta^\perp
\right)
\right|^2
\,ds.
\label{eq:app-radon-fourier-energy}
\end{equation}
By the directional assumption on the local edge-displacement mode
$v_{\xi_0,\rho_0}$ introduced above, its leading Fourier energy is
concentrated along the normal direction $\rho_0$. Therefore, the radial
Fourier line in \eqref{eq:app-radon-fourier-energy} captures the maximal
leading-order energy when
\begin{equation}
n_{\theta^\ast}^\perp\parallel\rho_0.
\end{equation}
Equivalently, the view-angle $\theta^\ast$ is tangent to the edge at
$\xi_0$. Combining this with \eqref{eq:app-radon-energy-objective} shows
that the tangent view maximizes the leading-order reduction in the
posterior variance of the localized directional quantity of interest.


%
%

\subsection{Local one-mode approximation for the star-shaped prior}
\label{app:star-shaped-local-mode}

We now justify the local one-mode assumption
\eqref{eq:app-local-one-mode} for the star-shaped parameterization introduced
in Section~\ref{ssec:prior}. The key observation is that, independently of
the particular parameter perturbation that generates it, the first-order
geometric deformation of a smooth boundary is determined locally by its
normal displacement.

Consider one star-shaped inclusion with center $C$ and radial function
\begin{equation}
B(\varphi)=\exp(Z(\varphi)),
\end{equation}
whose boundary is parameterized by
\begin{equation}
\gamma(\varphi)
=
C+B(\varphi)e_r(\varphi),
\qquad
e_r(\varphi)
=
\begin{pmatrix}
\cos\varphi\\
\sin\varphi
\end{pmatrix}.
\label{eq:app-star-boundary}
\end{equation}
At each boundary point, any first-order perturbation
$\delta\gamma(\varphi)\in\mathbb R^2$ admits the decomposition
\begin{equation}
\delta\gamma(\varphi)
=
V_n(\varphi)n(\varphi)
+
V_t(\varphi)t(\varphi),
\label{eq:app-normal-tangential-decomposition}
\end{equation}
where $n$ and $t$ denote the unit normal and tangent to the boundary,
respectively. The tangential component corresponds to a first-order
reparameterization of the boundary, whereas its geometric deformation as
a set is determined by the normal displacement $V_n$. Thus, perturbations
arising from radial, translational, or rotational uncertainty are all
represented locally through their corresponding normal displacement.

We now specialize to the shape perturbations induced by the latent radial
field $Z$. A perturbation $\delta Z$ gives
\begin{equation}
\delta B(\varphi)
=
B(\varphi)\delta Z(\varphi),
\end{equation}
and therefore
\begin{equation}
\delta\gamma(\varphi)
=
B(\varphi)\delta Z(\varphi)e_r(\varphi).
\label{eq:app-star-perturbation}
\end{equation}
Introducing
\begin{equation}
e_\varphi(\varphi)
=
\begin{pmatrix}
-\sin\varphi\\
\cos\varphi
\end{pmatrix},
\end{equation}
the boundary tangent satisfies
\begin{equation}
\gamma'(\varphi)
=
B'(\varphi)e_r(\varphi)
+
B(\varphi)e_\varphi(\varphi),
\end{equation}
and an outward unit normal is
\begin{equation}
n(\varphi)
=
\frac{
B(\varphi)e_r(\varphi)
-
B'(\varphi)e_\varphi(\varphi)
}{
\sqrt{B(\varphi)^2+B'(\varphi)^2}
}.
\label{eq:app-star-normal}
\end{equation}
Consequently, the normal component of the perturbation
\eqref{eq:app-star-perturbation} is
\begin{equation}
V_n(\varphi)
=
\delta\gamma(\varphi)^\top n(\varphi)
=
\frac{
B(\varphi)^2
}{
\sqrt{B(\varphi)^2+B'(\varphi)^2}
}
\delta Z(\varphi).
\label{eq:app-star-normal-displacement}
\end{equation}

Let $\xi_0=\gamma(\varphi_0)$ and restrict attention to the boundary segment
corresponding to $|\varphi-\varphi_0|\leq\varepsilon$. We choose the
regularity of the Matérn prior for $Z$ sufficiently high so that the
resulting radial perturbations are almost surely differentiable in
$\varphi$. Hence,
\begin{equation}
\delta Z(\varphi)
=
\delta Z(\varphi_0)
+
\delta Z'(\varphi_0)(\varphi-\varphi_0)
+
o(|\varphi-\varphi_0|).
\label{eq:app-local-Z-expansion}
\end{equation}
Defining the local random amplitude
\begin{equation}
A_{\xi_0}
:=
\delta Z(\varphi_0),
\end{equation}
we obtain
\begin{equation}
\delta Z(\varphi)
=
A_{\xi_0}
+
O(\varepsilon),
\qquad
|\varphi-\varphi_0|\leq\varepsilon.
\label{eq:app-local-radial-amplitude}
\end{equation}
Since the geometric factor multiplying $\delta Z(\varphi)$ in
\eqref{eq:app-star-normal-displacement} is smooth near $\varphi_0$, it
follows that
\begin{equation}
V_n(\varphi)
=
A_{\xi_0}
\frac{
B(\varphi_0)^2
}{
\sqrt{B(\varphi_0)^2+B'(\varphi_0)^2}
}
+
O(\varepsilon).
\label{eq:app-local-normal-amplitude}
\end{equation}
Thus, in a sufficiently small neighborhood of $\xi_0$, the first-order
geometric deformation of the boundary is governed by a single random
amplitude $A_{\xi_0}$.

It remains to transfer this local boundary deformation to image space.
Recall that the attenuation field is obtained through the push-forward map
\begin{equation}
X=F(C,Z).
\end{equation}
Let $(\bar C,\bar Z)$ denote a posterior reference parameter and
$\bar X=F(\bar C,\bar Z)$. For sufficiently small parameter perturbations,
the differentiability of $F$ gives
\begin{equation}
F(\bar C,\bar Z+\delta Z)
=
F(\bar C,\bar Z)
+
D_ZF(\bar C,\bar Z)[\delta Z]
+
r_F,
\label{eq:app-star-pushforward-linearization}
\end{equation}
where $r_F$ denotes the higher-order remainder. Combining the local
one-amplitude representation \eqref{eq:app-local-radial-amplitude} with
the linearity of $D_ZF(\bar C,\bar Z)$ shows that, after localization near
$\xi_0$, the induced image perturbation is governed to leading order by
the same scalar amplitude. After centering with respect to the current
posterior mean, we therefore obtain
\begin{equation}
M_{\phi_{\xi_0}}
\left(
X-m_X^{\mathrm{post}}
\right)
=
\widetilde A_{\xi_0}
v_{\xi_0,\rho_0}
+
r_\varepsilon,
\label{eq:app-star-local-one-mode}
\end{equation}
where
\begin{equation}
\widetilde A_{\xi_0}
=
A_{\xi_0}
-
\mathbb E
\left[
A_{\xi_0}
\mid
\boldsymbol Y_\Theta=\boldsymbol y_\Theta
\right],
\end{equation}
$v_{\xi_0,\rho_0}\in L^2(\Omega)$ denotes the localized image-space mode
induced by the corresponding normal displacement of the edge, and
$r_\varepsilon$ collects the higher-order and non-dominant local
perturbations.

Provided these remainder terms vanish in conditional mean square as the
localization neighborhood shrinks,
\begin{equation}
\mathbb E
\left[
\|r_\varepsilon\|_{L^2(\Omega)}^2
\mid
\boldsymbol Y_\Theta=\boldsymbol y_\Theta
\right]
\rightarrow 0,
\qquad
\varepsilon\rightarrow 0,
\end{equation}
equation~\eqref{eq:app-star-local-one-mode} establishes the local one-mode
property assumed in \eqref{eq:app-local-one-mode}. Hence, to leading order,
the localized posterior uncertainty generated by the star-shaped
push-forward prior is described by a single image-space mode associated
with normal displacement of the smooth edge through $\xi_0$.

Finally, the local normal-displacement mode also has the directional
Fourier structure assumed in the main argument. Let $t_0\perp\rho_0$
denote the tangent direction at $\xi_0$, and introduce the local
coordinates
\begin{equation}
s=(\xi-\xi_0)^\top\rho_0,
\qquad
\tau=(\xi-\xi_0)^\top t_0.
\end{equation}
Since the boundary is smooth, it is locally approximated by its tangent
line, and the image perturbation induced by a normal displacement varies
primarily across the edge. Consequently, to leading order the localized
mode has the form
\begin{equation}
v_{\xi_0,\rho_0}(\xi)
\approx
a(\tau)h(s),
\label{eq:app-local-mode-separable}
\end{equation}
where $h$ describes the variation across the edge and $a$ is a smooth
localization along the edge. Its Fourier transform therefore satisfies
\begin{equation}
\widehat v_{\xi_0,\rho_0}
(\omega_\rho\rho_0+\omega_t t_0)
\approx
\widehat h(\omega_\rho)\widehat a(\omega_t).
\label{eq:app-local-mode-fourier}
\end{equation}
Since $a$ varies smoothly in the tangential direction, its Fourier
transform is concentrated near $\omega_t=0$. Hence the leading
directional Fourier content of $v_{\xi_0,\rho_0}$ is concentrated along
the normal direction $\rho_0$, which verifies the directional assumption
used in the main proof.

\section*{Acknowledgments} We thank Yiqiu Dong and Jakob Lemvig for valuable discussions, and Sebastian Krumscheid and Truong Vinh Hoang for introducing us to the PACE method.

\section*{Funding} M.G.S. and H.Y. are supported by the Finnish Ministry of Education and Culture's Pilot for Doctoral Programmes (Pilot project Mathematics of Sensing, Imaging and Modelling, DREAM). H.Y. is also supported by the Research Council of Finland (RCF) through the Flagship of Advanced Mathematics for Sensing, Imaging and Modelling (FAME; grant number 359186). B.M.A. is partially supported by the RCF under grant numbers 371523, 359182, and 359186.

\section*{Author contributions}M.G.S.: Conceptualization, Methodology, Algorithm development, Investigation, Software, Formal analysis, Visualization, Writing -- original draft, Writing -- review \& editing. H.Y.: Methodology, Software, Algorithm development, Writing -- review \& editing. B.M.A.: Methodology, Software, Algorithm development, Supervision, Funding acquisition, Writing -- review \& editing.

$\,$

$\,$

\end{document}